\documentclass[preprintnumbers, showpacs, floatfix,onecolumn,preprintnumbers, letterpaper, superscriptaddress,nofootinbib]{revtex4}
\usepackage{amsfonts}
\usepackage{amsmath}
\usepackage{amssymb,epsf}
\usepackage{latexsym}
\usepackage{graphicx,epsfig}
\usepackage{amssymb}
\usepackage{float}
\usepackage{subfigure}
\usepackage{epstopdf}
\usepackage[colorlinks=true,citecolor=blue,linkcolor=blue,urlcolor=black]{hyperref}
\usepackage{dcolumn}
\usepackage{psfrag}
\usepackage{wrapfig}
\usepackage{makeidx}
\usepackage{epsf}

\begin{document}

\title{Lifshitz scaling effects on the holographic $p$-wave superconductors coupled to nonlinear electrodynamics}
\author{Mahya Mohammadi}
\email{mahya689mohammadi@gmail.com}
\affiliation{Physics Department and Biruni Observatory, Shiraz University, Shiraz 71454,
Iran}
\author{Ahmad Sheykhi}
\email{asheykhi@shirazu.ac.ir}
\affiliation{Physics Department and Biruni Observatory, Shiraz University, Shiraz 71454,
Iran}
\affiliation{Research Institute for Astronomy and Astrophysics of Maragha (RIAAM), P.O. Box 55134-441,
Maragha, Iran}
\affiliation{Institut f\"{u}r Physik, Universit\"{a}t Oldenburg,
Postfach 2503 D-26111 Oldenburg, Germany}

\begin{abstract}
We employ gauge/gravity duality to study the effects of Lifshitz
scaling on the holographic $p$-wave superconductors in the
presence of Born-Infeld (BI) nonlinear electrodynamics.
By using the shooting method in the probe limit, we
calculate the relation between critical temperature $T_{c}$ and
$\rho^{z/d}$ numerically for different values of mass, nonlinear
parameter $b$ and Lifshitz critical exponent $z$ in various
dimensions. We observe that critical temperature decreases by
increasing $b$, $z$ or the mass parameter $m$ which makes
conductor/superconductor phase transition harder to form.
In addition, we analyze the electrical conductivity and find the
behavior of real and imaginary parts as a function of frequency
which depend on the model parameters. However, some universal
behaviors are seen. For instance at low frequencies, real part of
conductivity shows a delta function behavior while the imaginary
part has a pole which means that these two parts are connected to
each other through Kramers-Kronig relation. The behavior of real
part of conductivity in the large frequency regime can be achieved
by $Re[\sigma]=\omega^{D-4}$. Furthermore, with increasing the
Lifshitz scaling $z$, the energy gap and the minimum values of the
real and imaginary parts become unclear.

\end{abstract}
\maketitle

\section{Introduction}
After the discovery of superconductivity, a considerable attempts
have been done to understand the different aspects of this
phenomenon \cite{Dahi}. The most successful way to describe
the superconductivity within a microscopic theory proposed by
Bardeen, Cooper and Schrieffer (BCS) who could address the
superconductivity as a microscopic effect originates from
condensation of Cooper pairs into a boson-like state \cite{BCS57}.
However, it wasn't prosperous to declare this effect completely.
More specifically, it is only practical for $s$-wave
superconductors while there are other types of superconductors
such as $p$-waves and $d$-waves \cite{superp,alkac}. Besides, the
BCS theory cannot explain the mechanism of high temperature
superconductors because the Cooper pairs are decoupled and no
longer exist when the temperature of the system becomes high
\cite{alkac}. By applying the AdS/CFT correspondence, which
relates the strong coupling conformal field theory on the boundary
in $d$-dimensions to a weak coupling gravity in
$(d+1)$-dimensional bulk, the novel idea of holographic
superconductors was proposed by Hartnoll et.
al.\cite{Maldacena,H08}. The holographic superconductors opened up
a new window for studying all kinds of superconductors. Based on
this idea, the system faces with spontaneous $U(1)$ symmetry
breaking during a phase transition from a black hole with no hair
(normal phase) to a hairy one (superconducting phase) below the
critical temperature \cite{H11}. The holographic
superconductors have many properties similar to real world
superconductors such as the second order phase transition at the
critical temperature. One of the most important success of
holographic superconductors is that all these models show an exact
mean-field behaviors at the critical temperature, just like the
Landau-Gindzburg (LG) theory for continuous phase transitions. All
the critical exponents for the order parameter at $T_{c} $ are
equal to $1/2$  \cite{H08}. Nevertheless, the idea of holographic
superconductor is a theoretical attempt for addressing the puzzle
of high temperature superconduction and the experimental features
of this idea is still an open problem. In the past decade,
holographic superconductors grabs a lot of attentions (see e.g.
\cite{H09,Hg09,H11,G98,W98,HR08,R10,Gu09,HHH08,JCH10,SSh16,SH16,cai15,SHsh(17),
Ge10,Ge12,Kuang13,Pan11,CAI11, SHSH(16),shSh(16),Doa, Afsoon,
cai10,yao13,n4,n5,n6,Gan1,mahya}).

While the $p$-wave and $d$-wave superconductors are known as strong
coupling superconductors, the investigations on them based on the
holographic setup attracts a lot of attentions during the past
years (see e.g. \cite{Caip,cai13p,Donos,Gubser,Roberts8,zeng11,
cai11p,pando12,momeni12p,gangopadhyay12,chaturverdip15,mahyap,mahyap1,francessco1,francessco2}).
There are different ways to investigate the properties of
holographic $p$-wave superconductors including condensation of a
real or complex massive charged vector field in the gravity side
which is dual to vector order parameter in the boundary as well as
introducing a 2-form field or a $SU(2)$ Yang-Mills gauge field in
the bulk as the origin of spin-1 order parameter
\cite{Caip,cai13p,Donos,Gubser,chaturverdip15}. The impact of
Gauss-Bonnet term on the holographic $p$-wave superconductor is
characterized in different works \cite{caipp,gaussp1,mahyagaussp}.
Moreover, in the framework of condensed matter physics, a
dynamical scaling appears near the critical point. In the vicinity
of critical point, the scale transformation turns to be \cite{Bu}:
\begin{equation}\label{eq00}
t \rightarrow \lambda^{z} t, \ \    x_{i} \rightarrow \lambda x_{i}, \  \  z\neq0.
\end{equation}
when $z=1$, the usual AdS spacetime is obtained.  Otherwise, the
temporal and the spatial coordinates scale anisotropically. Many
investigations have been done regarding the anisotropic
superconductors \cite{smylie}. Nowadays, the applications of
superconductors aren't limited to physics. However, the actual use
of superconducting devices is limited due to the fact that they
must be cooled to low temperatures to become superconductors and
hence cannot use for ordinary (lab) temperatures. So by passing
time, $p$-wave and $d$-wave superconductors as candidates for high
temperature superconductors grab a lot of attentions. In some
cases $p$-wave superconductors show anisotropic behavior
\cite{superp}. In addition, heavy fermion compounds and other
materials including high $T_{c} $ superconductors have a metallic
phase (dubbed as strange metal) whose properties cannot be
explained within the ordinary Landau-Fermi liquid theory. In this
phase some quantities exhibit universal behavior such as the
resistivity, which is linear in the temperature $\sigma \sim T$.
Such universal properties are believed to be the consequence of
quantum criticality. At the quantum critical point there is a
Lifshitz scaling same as Eq.  (\ref{eq00})\cite{carlos}. Many
researches carried out in Lifshitz scaling by using the
holographic approach (see e.g.
\cite{LU,Natsuume,Sherkatghanad,Zhao14}). However, all the
previous works about the effects of Lifshitz scaling on
holographic $p$-wave superconductor were done by considering a
SU(2) Yang-Mills gauge field in the bulk in the presence of Maxwell electrodynamics \cite{Bu,LU}. In this
work, we have explored the effects of Lifshitz scaling on
holographic $p$-wave superconductor by introducing a charged
vector field in the bulk. As the consequence of this method all
following calculations to analyze the behavior of condensation and
conductivity became different from the previous works. However,
there are good agreements between our results with the case of
$m^{2}=0$ in \cite{LU}. This approach allows us to consider the
effects of mass term as well as the spacetime dimension and
Lifshitz scaling $z$, while in the previous method the mass term
plays no role and sets to zero. Besides, we can study the effect
of conductivity in easier way by turning on only the component $\delta A_{y}=A_{y} e^{-i \omega t}$ as perturbation on the black hole
background in compared to \cite{LU}.
Moreover,
It is worthful to investigate this model
in the presence of nonlinear electrodynamics \cite{Bu,LU}. There
exist several types of nonlinear electrodynamics in the
literatures, including BI \cite{25}, Exponential \cite{hendi},
Logarithmic \cite{log} and Power-Maxwell \cite{SSh16}. Among them,
the BI nonlinear electrodynamics is perhaps the most well-known
which was proposed to address the problem of the divergence of the
electrical field at the position of a point particle. It was later
pointed out that BI Lagrangian could be reproduced through string
theory, and its action naturally possesses electric-magnetic
duality invariance which makes it suitable for describing gauge
fields arising from open strings on D-branes
\cite{25,hendi,mahya,mahyap1}. Disclosing the effects of the
nonlinear electrodynamics on the behavior of superconductors is of
interest both for practical applications and for the study of the
fundamental properties of the materials. In any practical
electronic application, the designer must know how much power the
conductors can handle and at which power level nonlinear effects
such as harmonic generation and intermodulation (IM) distortion
become appreciable. Therefore, the magnitude and the detailed
nature of the nonlinear effects must be measured and understood in
order to facilitate widespread application of superconductors in
microwave frequency electronics. The effects of nonlinear
electrodynamics widely studied in the literatures (see e.g
\cite{cchin,rose,Yampol,Coffey, Jeff,xia}). Although  these
investigations extend over many years, new interest in these
nonlinear effects has been kindled since the discovery of the
high-T, oxide materials \cite{xia}. Therefore, it is worthy to
provide theoretical approach for prediction of the behavior of
real high temperatures superconductors in the presence of
nonlinear electrodynamics based on the holographic approach. For
example we find out that by increasing the effect of nonlinearity,
critical temperature decreases. The effects of nonlinear
electrodynamics on the holographic superconductors have been
explored widely in the literatures (see e.g.
\cite{n4,SH16,SSh16,SHsh(17),SHSH(16),shSh(16),n5,n6,mahya,mahyap1}).

Our aim in this work is to investigate the effects of Lifshitz
scaling on the holographic $p$-wave superconductor in arbitrary
dimensions. We shall study the phase transition between conductor
and superconductor which depends crucially on the parameters $m$,
$b$ and $z$. In spite of the fact that there is a dynamical
exponent, we find out that the condensation has mean-field
behavior near the critical temperature which is the same as AdS
spacetime. Additionally, the electrical conductivity in
gauge/gravity correspondence is achieved by imposing appropriate
perturbation on the gauge field. Besides, the conductivity
formula, we calculate the behavior of both real and imaginary
parts of conductivity as a function of frequency. Although, the
obvious differences in graphs based on our choice of $m$, $b$ and
$z$, they follow same trends in some cases. A good illustration of
this is obeying the Kramers-kronig relation by having a delta
function and a pole in real and imaginary parts of conductivity.
However, the gap frequency which is occurred below the critical
temperature, becomes less obvious by enlarging the anisotropy
between space and time. The effects of nonlinearity
parameter on the conductivity will be clearly indicated via
graphs. Our choice of mass in each dimension has a direct outcome
on the effect of BI nonlinear electrodynamics on conductivity but
generally the gap energy and minimum of conductivity shift toward
larger values of frequency by enlarging the nonlinearity effects.

This article is organized as follows. In section \ref{section1},
we analyze the holographic setup via condensation of the vector
field in the context of Lifshitz spacetime and in the presence of
 nonlinear  BI electrodynamics.
 We explore the electrical conductivity of this model in section \ref{section2}. Finally, our outcomes are
summarized in section \ref{section3}.

\section{Holographic $p$-wave superconductors with Lifshitz scaling}\label{section1}
\begin{figure*}[t]
\centering
\subfigure[~$z=1$, $m^{2}=0$]{\includegraphics[width=0.4\textwidth]{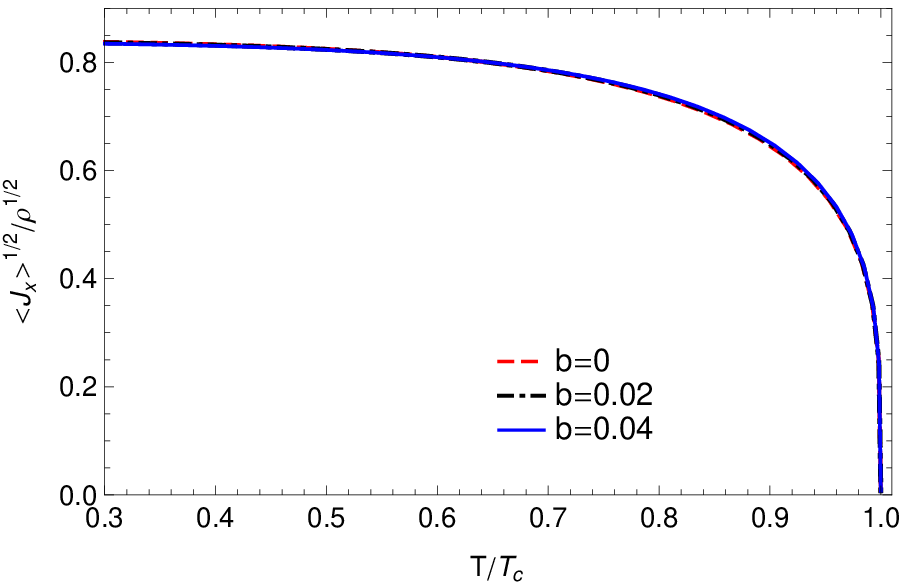}} \qquad %
\subfigure[~$z=2$, $m^{2}=0$]{\includegraphics[width=0.4\textwidth]{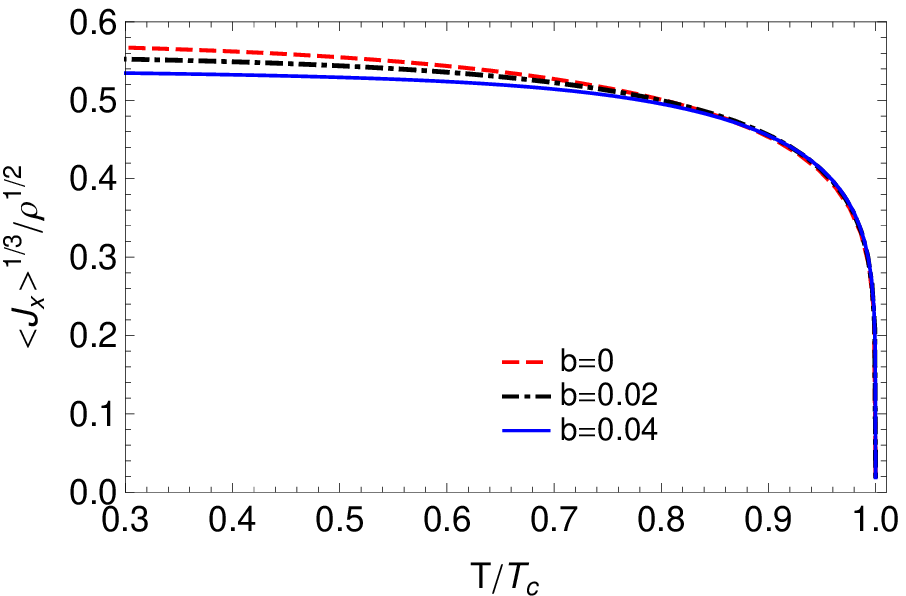}} \qquad %
\subfigure[~$z=1$, $m^{2}=3/4$]{\includegraphics[width=0.4\textwidth]{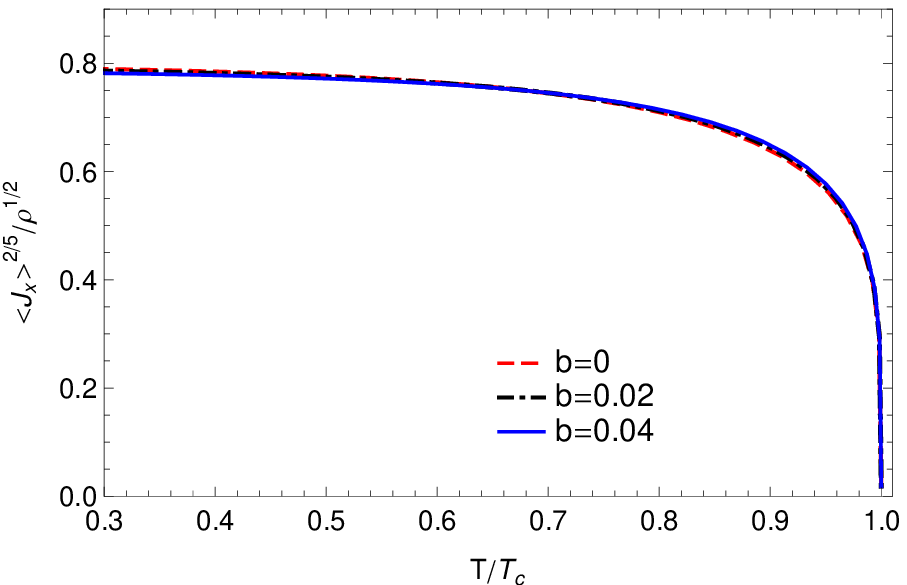}} \qquad %
\subfigure[~$z=2$, $m^{2}=-3/4$]{\includegraphics[width=0.4\textwidth]{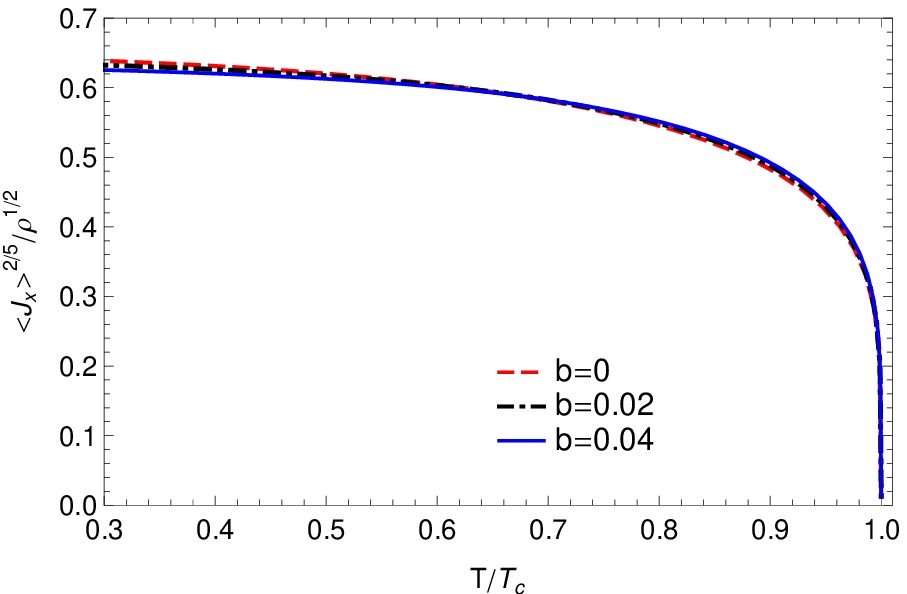}} \qquad %
\caption{The behavior of the condensation parameter as a function
of temperature for different values of model parameters in
$D=d+2=4$ dimension.} \label{fig1}
\end{figure*}

\begin{figure*}[t]
\centering
\subfigure[~$z=1$, $m^{2}=0$]{\includegraphics[width=0.4\textwidth]{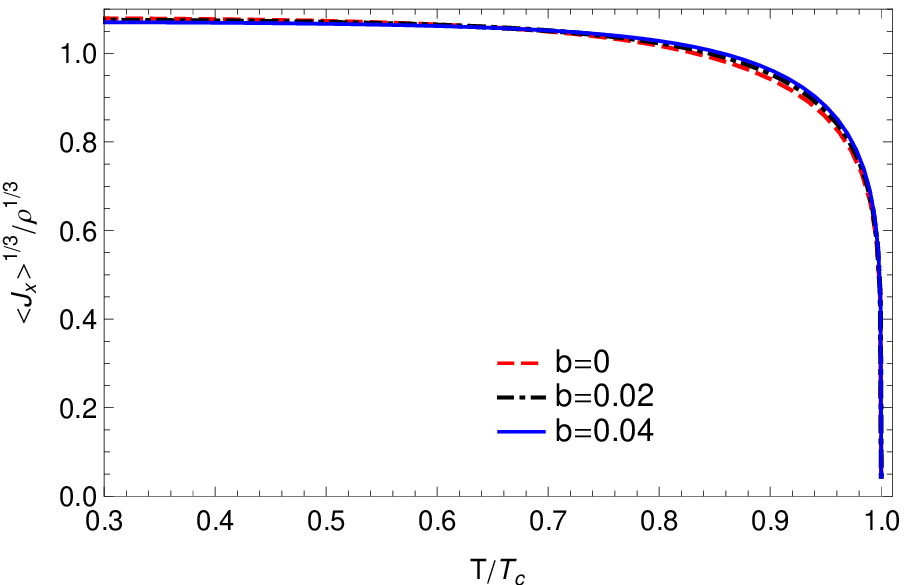}} \qquad %
\subfigure[~$z=2$, $m^{2}=0$]{\includegraphics[width=0.4\textwidth]{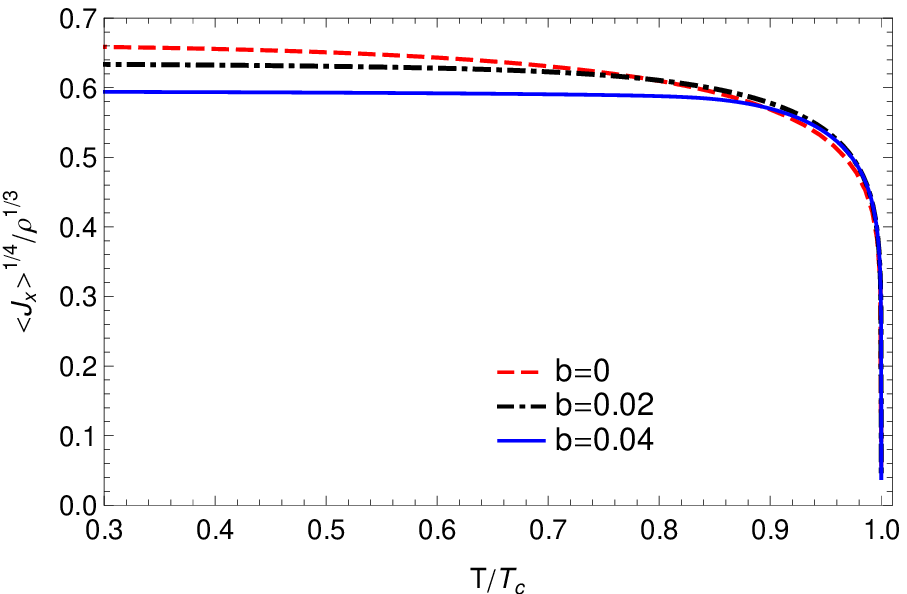}} \qquad %
\subfigure[~$z=1$, $m^{2}=-3/4$]{\includegraphics[width=0.4\textwidth]{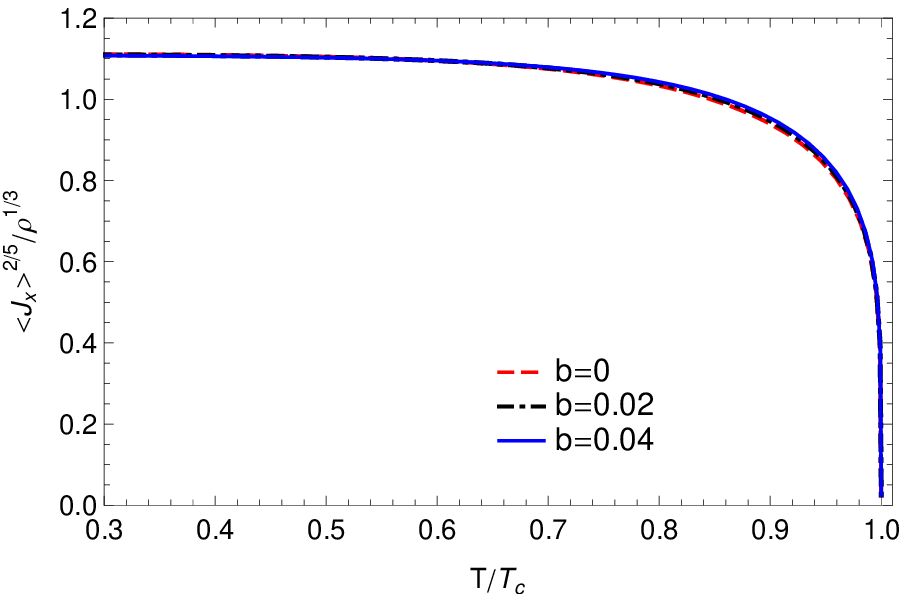}} \qquad %
\subfigure[~$z=2$, $m^{2}=-2$]{\includegraphics[width=0.4\textwidth]{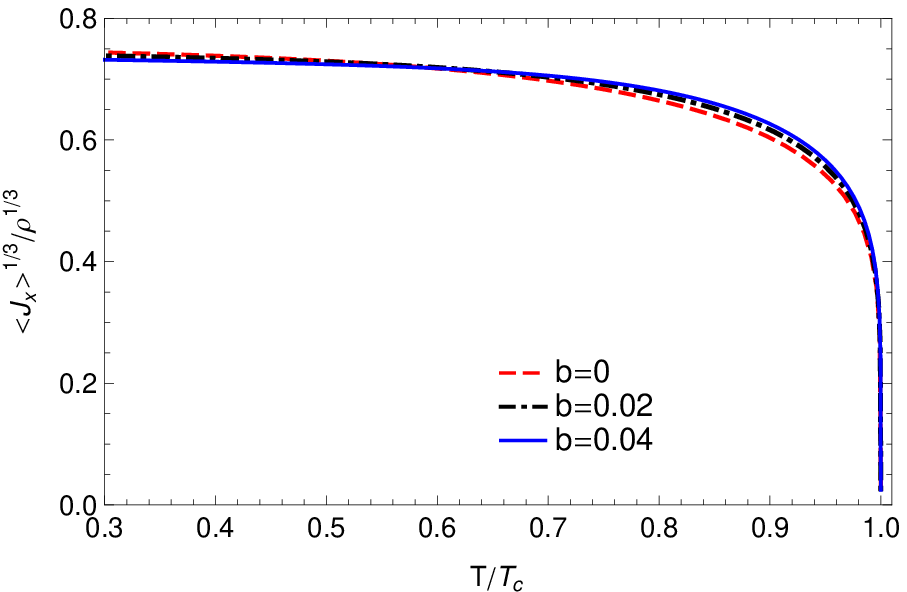}} \qquad %
\caption{The behavior of the condensation parameter as a function
of the temperature for different values of the mass, the dynamical
critical exponent and nonlinearity parameters in $D=d+2=5$
dimension.} \label{fig2}
\end{figure*}

We consider a $(d+1)$-dimensional holographic $p$-wave
superconductor living on the boundary of a $(d+2)$-dimensional
Lifshitz black hole in the presence of BI nonlinear
electrodynamics, which is described by the following action
\cite{cai13p}
\begin{eqnarray}
&&S =\int d^{d+2}x\sqrt{-g} \left(R-2 \Lambda+\mathcal{L}_{m}\right), \notag \\
&& \mathcal{L}_{m}=
\mathcal{L}(\mathcal{F})-\frac{1}{2}\rho_{\mu\nu}^{\dagger}
\rho^{\mu\nu}-m^{2} \rho_{\mu}^{\dagger} \rho^{\mu} + i q \gamma
\rho_{\mu} \rho_{\nu}^{\dagger} F^{\mu\nu} ,\label{act}
\end{eqnarray}%
where $m$ and $q$ are the mass and charge of vector field
$\rho_{\mu}$. The metric determinant, the Ricci scalar and the
negative cosmological constant are demonstrated by $g$, $R$ and
$\Lambda$, respectively. In terms of the radius of Lifshitz
spacetime, $l$, we can formulate the cosmological constant as
\cite{LU}
\begin{equation}
\Lambda=-\frac{(d+z-1)(d+z)}{2 l^2},
\end{equation}
where $z$, the Lifshitz scaling, is a dynamical critical exponent
standing for the anisotropy between space and time. Hereafter, for
simplicity we set $l=1$. The Lagrangian density of the BI
nonlinear electrodynamics $\mathcal{L}(\mathcal{F})$ is given by
\cite{25}
\begin{equation}
\mathcal{L}(\mathcal{F})=\frac{1}{b}\left( 1-\sqrt{1+\frac{b\mathcal{F}}{2}}%
\right),
\end{equation}%
where $\mathcal{F}=F_{\mu\nu} F^{\mu\nu}$ is the Maxwell invariant
and $b$, with dimension of $[{length}]^2$, represents the strength
of nonlinearity. When $b\rightarrow0$, $\mathcal{L}(\mathcal{F})$
restores the standard Maxwell Lagrangian
$\mathcal{L}_{max}(\mathcal{F})=-\mathcal{F}/4$. The strength of
the electromagnetic field is represented by
$F_{\mu\nu}=\nabla_{\mu} A_{\nu}-\nabla_{\nu} A_{\mu}$ where
$A_{\mu}$ is the vector potential. In the Lagrangian of the matter
field $\mathcal{L}_{m}$, by using the covariant derivative
$D_{\mu}=\nabla_{\mu}- i q A_{\mu}$, we can set
$\rho_{\mu\nu}=D_{\mu} \rho_{\nu}-D_{\nu} \rho_{\mu}$. The last
term in the matter Lagrangian shows the strength of interaction
between $\rho_{\mu}$ and $A_{\mu}$ with $\gamma$ as the magnetic
moment in the case with an applied magnetic field which will be
ignored in this work.  The equations of motion, for the matter
fields, can be obtained by varying the action (\ref{act}) with
respect to the gauge field $A_{\mu}$ and the vector field
$\rho_{\mu}$,
\begin{equation}\label{eqmax}
\nabla ^{\nu }\left[-4 \mathcal{L}_{\mathcal{F}}  F_{\nu \mu
}\right] =i q \left(\rho ^{\nu } \rho ^{\dagger }{}_{\nu \mu }-
\rho ^{\nu \dagger } \rho _{\nu \mu }\right)+ i q \gamma
\nabla^{\nu} \left(\rho _{\nu } \rho ^{\dagger }{}_{\mu }-\rho
^{\dagger }{}_{\nu } \rho _{\mu } \right),
\end{equation}
\begin{equation}\label{eqvector}
D ^{\nu } \rho_{\nu \mu }-m^{2} \rho_{\mu }+ i q \gamma \rho^{\nu
} F_{\nu\mu} =0,
\end{equation}
where $\mathcal{L}_{\mathcal{F}}=\partial
\mathcal{L(\mathcal{F})}/\partial{\mathcal{F}}$. Since we work in
the probe limit, the background spacetime is not affected by the
vector and gauge fields. Thus, we can write down the metric of
$(d+2)$-dimensional Lifshitz spacetime as \cite{LU}
\begin{eqnarray} \label{metric}
&&{ds}^{2}=-r^{2 z} f(r) {dt}^{2}+\frac{{dr}^{2}}{r^{2} f(r)}+r^{2} \sum _{i=1}^{d}{dx_{i}}^{2}%
,\\
&&f(r)=1-\frac{r_{+}^{d+z}}{r^{d+z}}.\label{eqf} %
\end{eqnarray}%
where $r_{+}$ denotes the black hole horizon obeying $f(r_{+})=0$.
We also assume the vector and gauge field has the following form
\begin{eqnarray}
&& \rho_{\nu} dx^{\nu}=\rho_x(r) dx, \ \ \ \ A_{\nu} dx^{\nu}=\phi
(r) dt. \label{rhoA}
\end{eqnarray}%
The regularity condition for the gauge field, on the horizon,
implies that $\phi(r_{+})=0$ \cite{Bu}. The Hawking temperature of
the black hole, associated with the horizon, is defined \cite{LU}
\begin{equation}
T=\frac{f^{^{\prime }}(r_{+})}{4\pi }=\frac{(d+z) r_{+}^{z}}{4\pi}. \label{temp}
\end{equation}%
Inserting Eqs. (\ref{metric}) and (\ref{rhoA}) in the field Eqs.
(\ref{eqmax}) and (\ref{eqvector}), we arrive at
\begin{equation}\label{eqphi}
\phi ''(r)+ \left[\frac{d-z+1}{r}-\frac{b d }{ r^{2 z-1}} \phi '^2(r)\right]
\phi '(r)-\frac{ 2 q^2 \rho_{x}^2(r)  }{f(r) r^{4}}\left[1-\frac{b}{r^{2 z -2}}
 \phi '^2(r)\right]^{3/2}\phi (r)=0,
\end{equation}
\begin{equation}\label{eqrho}
\rho _x''(r)+\left[\frac{d+z-1}{r}+\frac{f'(r)}{f(r)}\right]\rho _x'(r)
+\frac{\rho _x(r) }{r^2 f(r)}\left[\frac{q^2 \phi^2 (r)}{f(r) r^{2 z}}-m^2\right]=0,
\end{equation}
where the prime denotes derivative with respect to $r$. The linear
electrodynamic form of the above equations of motion are recovered
in the limiting case where $b\rightarrow0$ \cite{LU}. In the
remaining part of this paper without loss of generality, we will
set $r_{+}=1$ and $q=1$. At the boundary where $r\rightarrow
\infty$, the above equations have the asymptotic solutions as
\begin{equation}
\phi (r)= \left\{
\begin{array}{lr}
\mu- \rho r^{z-d}+\cdots, & z<d\\
\bigskip\\
\mu-\rho \log (r)+\cdots, & z=d\\
\end{array} \right.
\end{equation}%
\begin{equation}
\rho _x(r)= \frac{\rho _{x_+}}{r^{\Delta _+}}+\frac{\rho _{x_-}}{r^{\Delta _-}}+\cdots, \  \
\  \   \Delta _\pm=\frac{1}{2} \left[(d+z-2)\pm\sqrt{(d+z-2)^2+4 m^2}\right].
\end{equation}
According to gauge/gravity duality $\mu$, $\rho$, $\rho_{x_+}$ and
$\rho_{x_-}$ play, respectively, the role of the chemical
potential, charge density, $x$-component of the vacuum expectation
value of the order parameter $\langle J_{x} \rangle$ and source.
Since we expect the spontaneous $U(1)$ symmetry breaking so we
impose the source free condition i.e. $\rho_{x_-}=0$. In addition,
we follow the Breitenlohner-Freedman (BF) bound for our choice of
the mass,
\begin{equation}
m^{2}\geqslant -\frac{(d+z-2)^2}{4}.
\end{equation}
\begin{table*}[t]
\label{tab1}
\begin{center}
\begin{tabular}{ccc|c|c|c|}
\cline{4-6} & & & \multicolumn{3}{ c|}{$T_{c}/\rho^{z/d}$}  \\
\cline{4-6} & & & $b=0$ & $b=0.02$ & $b=0.04$  \\ \cline{1-6}
\multicolumn{1}{|c|}{$d=2$} &\multicolumn{1}{|c|}{$z=1$}
&\multicolumn{1}{|c|}{$m^{2}=0$} & 0.125 & 0.122 & 0.118  \\
\cline{1-6} \multicolumn{1}{|c|}{$d=2$}
&\multicolumn{1}{|c|}{$z=1$} &\multicolumn{1}{|c|}{$m^{2}=3/4$} &
0.102 & 0.096 & 0.089  \\ \cline{1-6} \multicolumn{1}{|c|}{$d=2$}
&\multicolumn{1}{|c|}{$z=2$} &\multicolumn{1}{|c|}{$m^{2}=0$} &
0.037 & 0.026 & 0.017  \\ \cline{1-6} \multicolumn{1}{|c|}{$d=2$}
&\multicolumn{1}{|c|}{$z=2$} &\multicolumn{1}{|c|}{$m^{2}=-3/4$} &
0.050 & 0.042 & 0.035  \\ \cline{1-6} \multicolumn{1}{|c|}{$d=3$}
&\multicolumn{1}{|c|}{$z=1$} &\multicolumn{1}{|c|}{$m^{2}=0$} &
0.200 & 0.184 & 0.169  \\ \cline{1-6} \multicolumn{1}{|c|}{$d=3$}
&\multicolumn{1}{|c|}{$z=1$} &\multicolumn{1}{|c|}{$m^{2}=-3/4$} &
0.224 & 0.215 & 0.207  \\ \cline{1-6} \multicolumn{1}{|c|}{$d=3$}
&\multicolumn{1}{|c|}{$z=2$} &\multicolumn{1}{|c|}{$m^{2}=0$} &
0.065 & 0.031 & 0.012  \\ \cline{1-6} \multicolumn{1}{|c|}{$d=3$}
&\multicolumn{1}{|c|}{$z=2$} &\multicolumn{1}{|c|}{$m^{2}=-2$} &
0.090 & 0.071 & 0.057  \\ \cline{1-6}
\end{tabular}
\caption{Numerical results for critical temperature $T_{c}$ for
different values of $z$, $m$ and $b$ in $D=d+2=4$ and $5$
spacetime dimesnion.}
\end{center}
\end{table*}
Considering the canonical ensemble with fixed $\rho$, and
employing the shooting method, we perform the numerical
calculations to derive the relation between critical temperature,
$T_{c}$, and charge density, $\rho^{z/d}$, for $z=1, 2$ in
$D=d+2=4$ and $5$ spacetime dimensions. Our results are summarized
in table I. We find out that increasing the values of $z$, $m$ and
nonlinearity $b$, in each dimension, hinders the superconductivity
phase by diminishing the critical temperature. Moreover, the
trends of condensation $\langle J_{x} \rangle^{1/(1+\Delta_{+})}$
as a function of temperature impressed by different values of $z$,
$m$ and $b$ are shown in figures \ref{fig1} and \ref{fig2}. Based
on these graphs, condensation goes down by raising $z$.

\newpage
\section{Electric conductivity}\label{section2}
\begin{figure*}[t]
\centering
\subfigure[~$z=1$, $m^{2}=0$, $b=0$]{\includegraphics[width=0.4\textwidth]{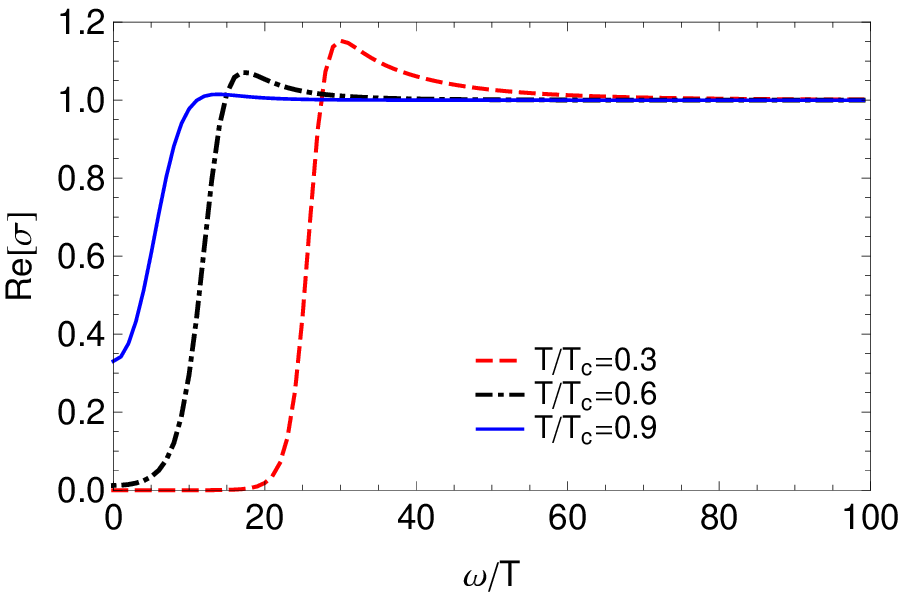}} \qquad %
\subfigure[~$z=1$, $m^{2}=0$, $b=0.04$]{\includegraphics[width=0.4\textwidth]{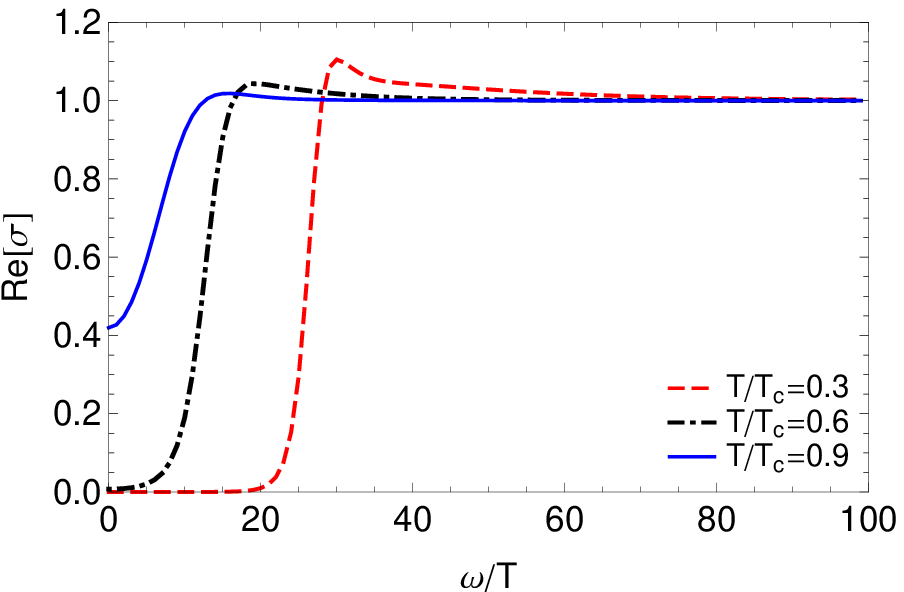}} \qquad %
\subfigure[~$z=2$, $m^{2}=0$, $b=0$]{\includegraphics[width=0.4\textwidth]{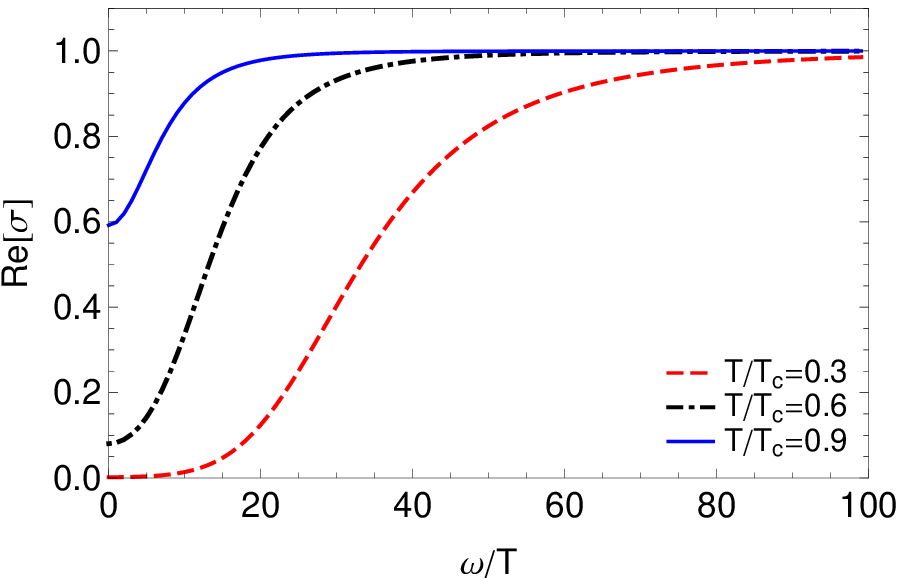}} \qquad %
\subfigure[~$z=2$, $m^{2}=0$, $b=0.04$]{\includegraphics[width=0.4\textwidth]{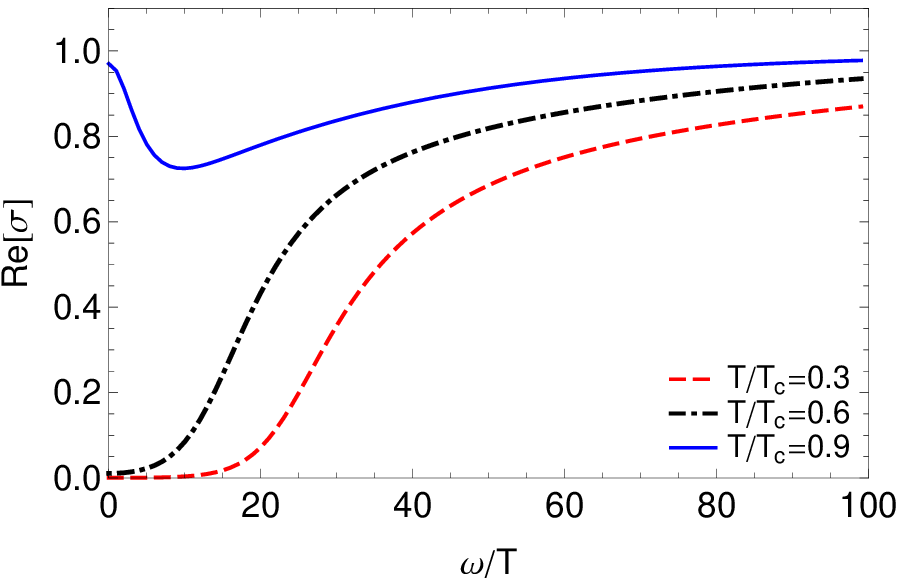}} \qquad %
\subfigure[~$z=1$, $m^{2}=3/4$, $b=0$]{\includegraphics[width=0.4\textwidth]{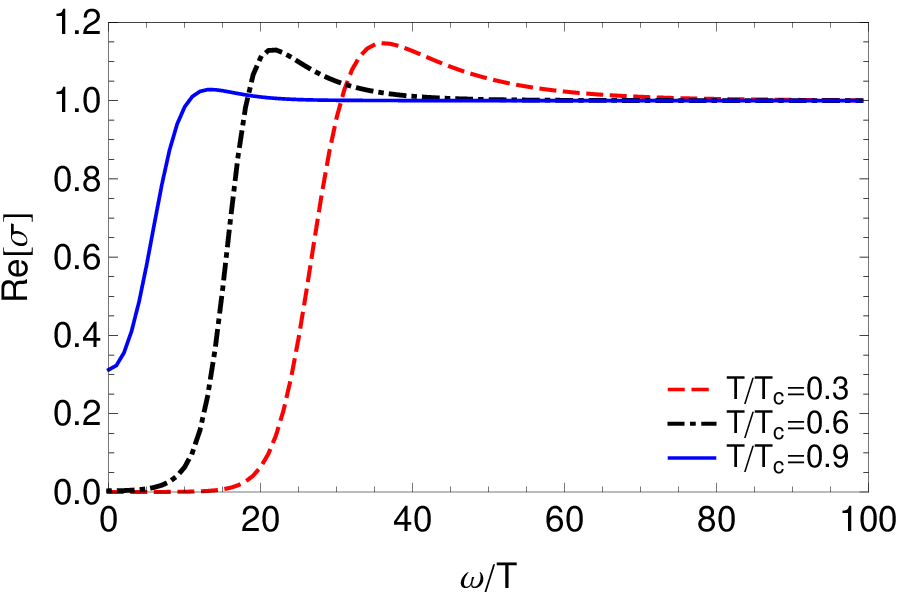}} \qquad %
\subfigure[~$z=1$, $m^{2}=3/4$, $b=0.04$]{\includegraphics[width=0.4\textwidth]{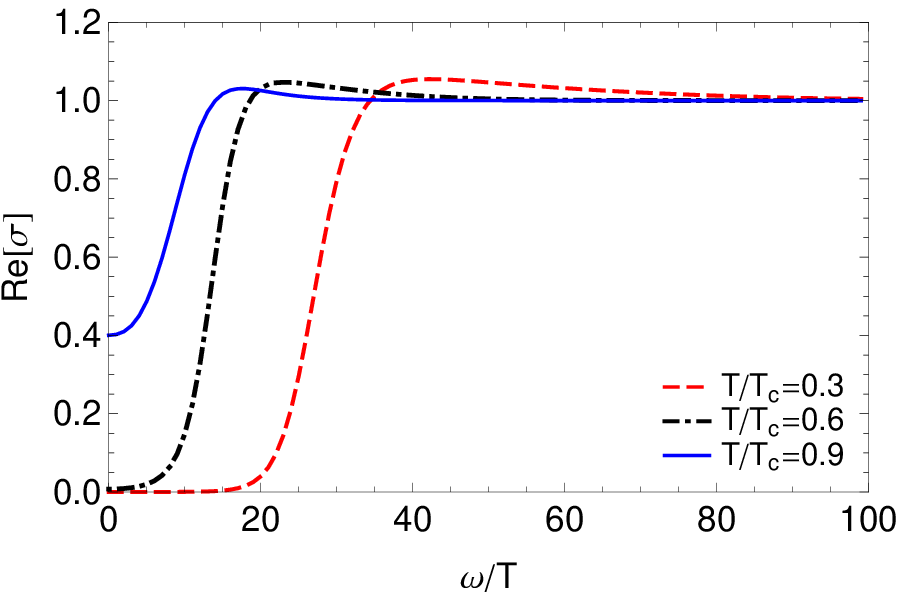}} \qquad %
\subfigure[~$z=2$, $m^{2}=-3/4$, $b=0$]{\includegraphics[width=0.4\textwidth]{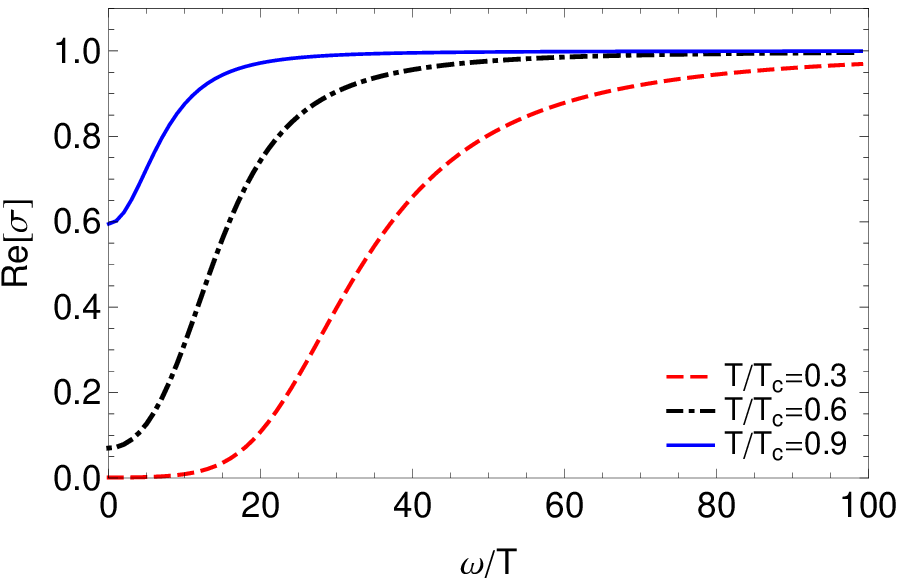}} \qquad %
\subfigure[~$z=2$, $m^{2}=-3/4$, $b=0.04$]{\includegraphics[width=0.4\textwidth]{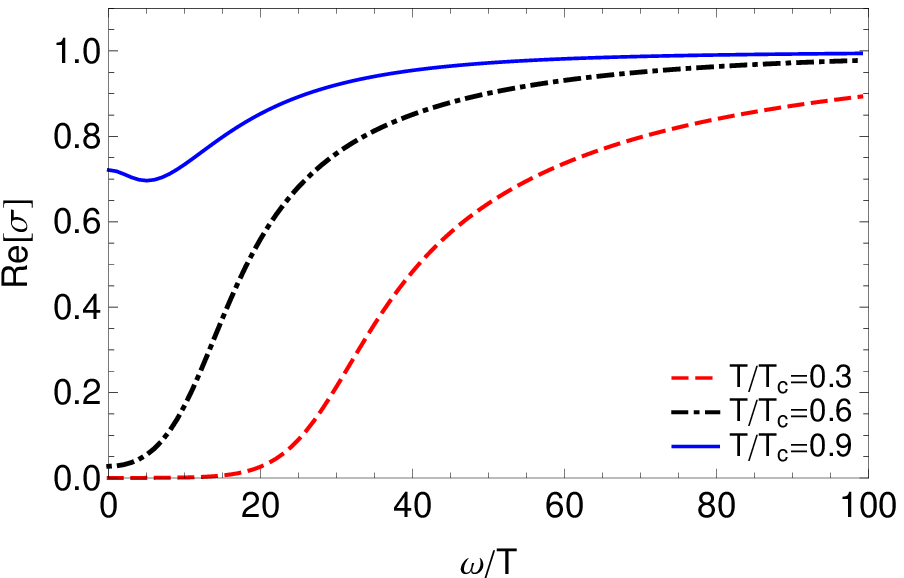}} \qquad %
\caption{The behavior of the real part of conductivity in $D=4$.}
\label{fig4}
\end{figure*}

\begin{figure*}[t]
\centering
\subfigure[~$z=1$, $m^{2}=0$, $b=0$]{\includegraphics[width=0.4\textwidth]{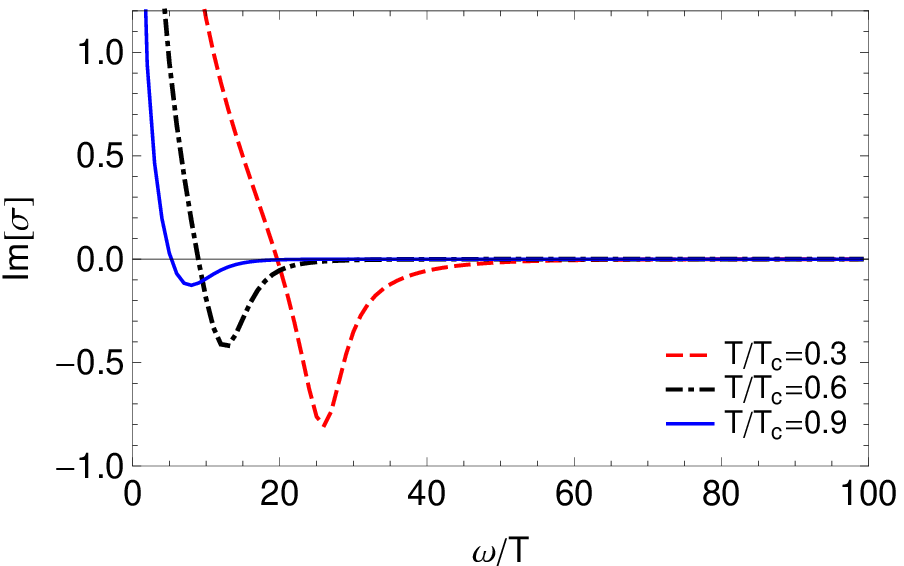}} \qquad %
\subfigure[~$z=1$, $m^{2}=0$, $b=0.04$]{\includegraphics[width=0.4\textwidth]{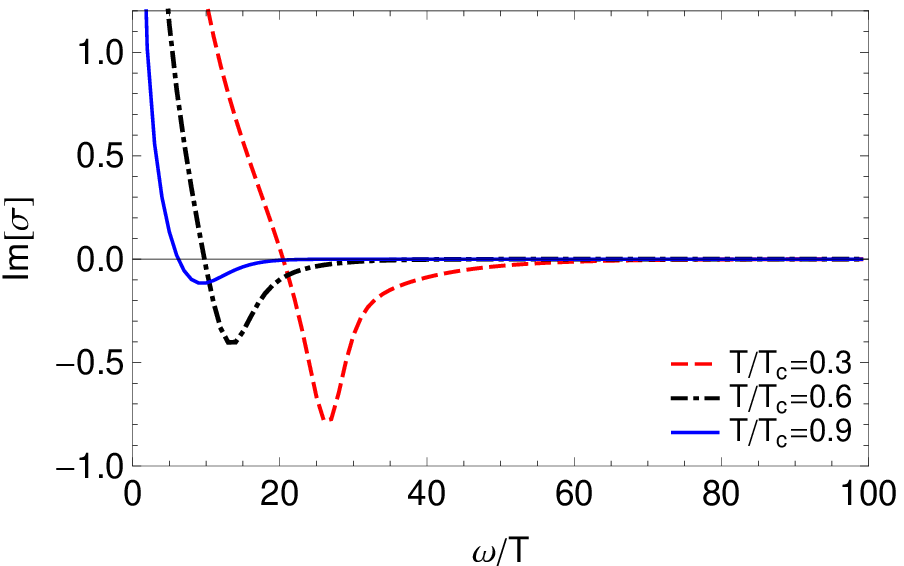}} \qquad %
\subfigure[~$z=2$, $m^{2}=0$, $b=0$]{\includegraphics[width=0.4\textwidth]{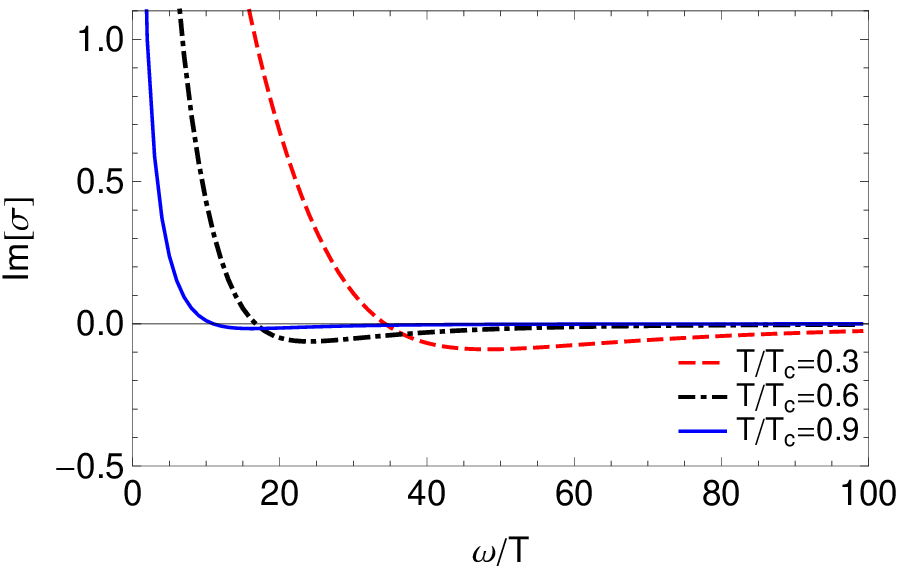}} \qquad %
\subfigure[~$z=2$, $m^{2}=0$, $b=0.04$]{\includegraphics[width=0.4\textwidth]{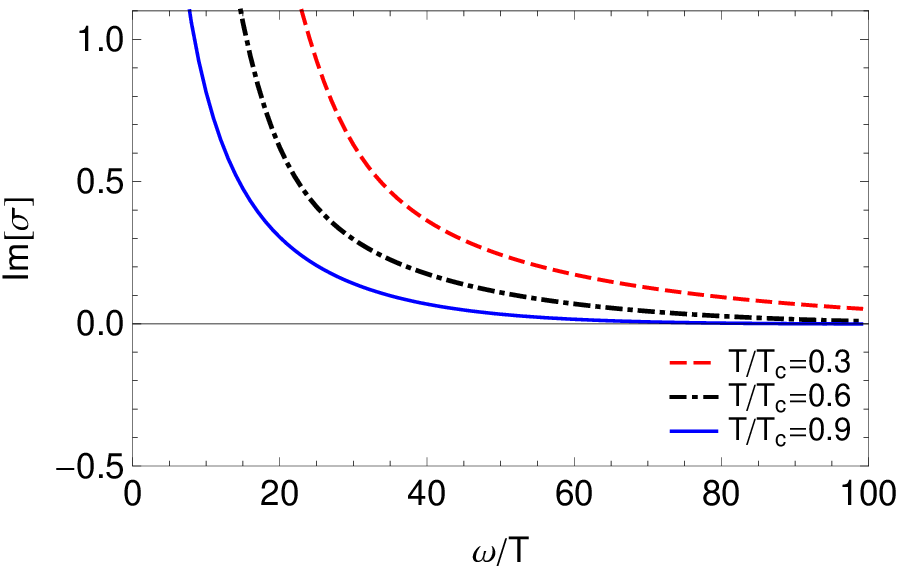}} \qquad %
\subfigure[~$z=1$, $m^{2}=3/4$, $b=0$]{\includegraphics[width=0.4\textwidth]{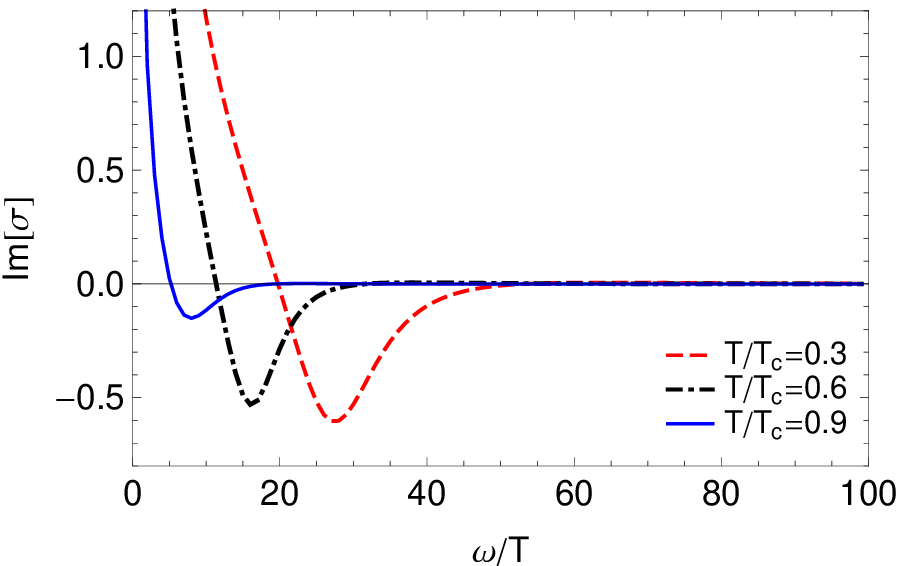}} \qquad %
\subfigure[~$z=1$, $m^{2}=3/4$, $b=0.04$]{\includegraphics[width=0.4\textwidth]{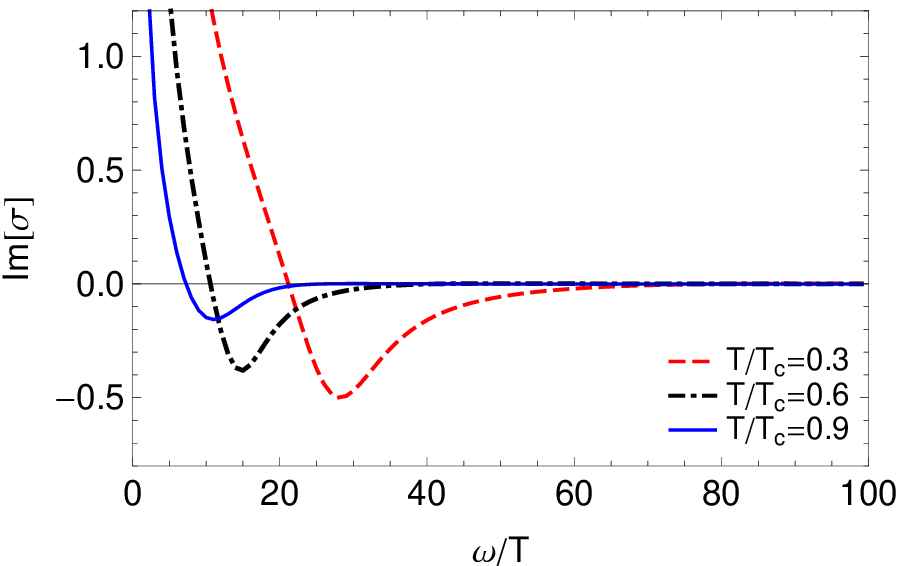}} \qquad %
\subfigure[~$z=2$, $m^{2}=-3/4$, $b=0$]{\includegraphics[width=0.4\textwidth]{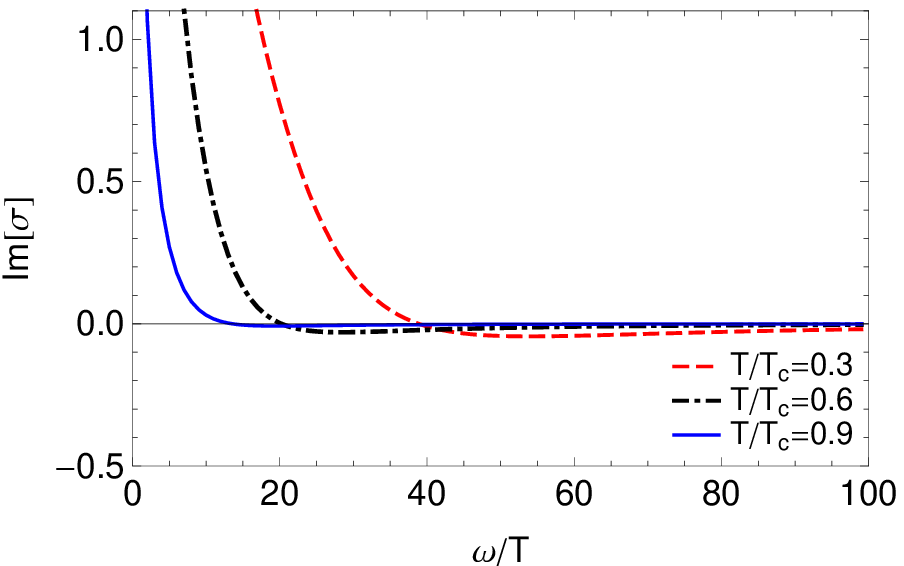}} \qquad %
\subfigure[~$z=2$, $m^{2}=-3/4$, $b=0.04$]{\includegraphics[width=0.4\textwidth]{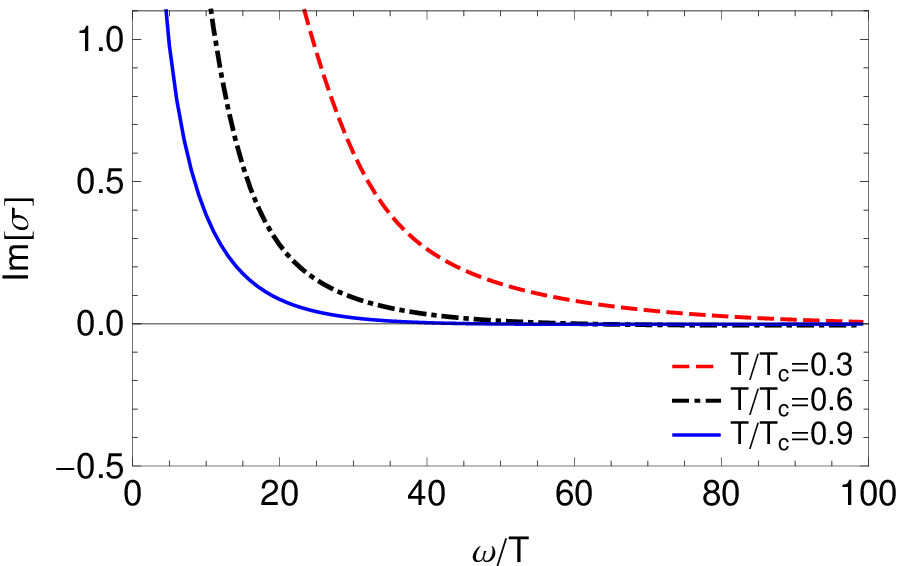}} \qquad %
\caption{The behavior of the imaginary part of conductivity in
$D=4$.} \label{fig5}
\end{figure*}

\begin{figure*}[t]
\centering
\subfigure[~$z=1$, $m^{2}=0$, $b=0$]{\includegraphics[width=0.4\textwidth]{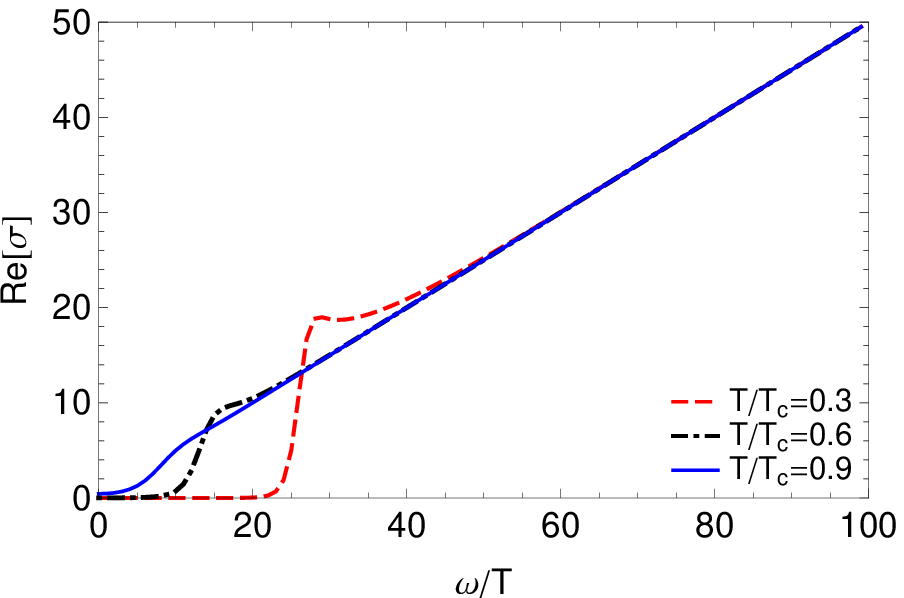}} \qquad %
\subfigure[~$z=1$, $m^{2}=0$, $b=0.04$]{\includegraphics[width=0.4\textwidth]{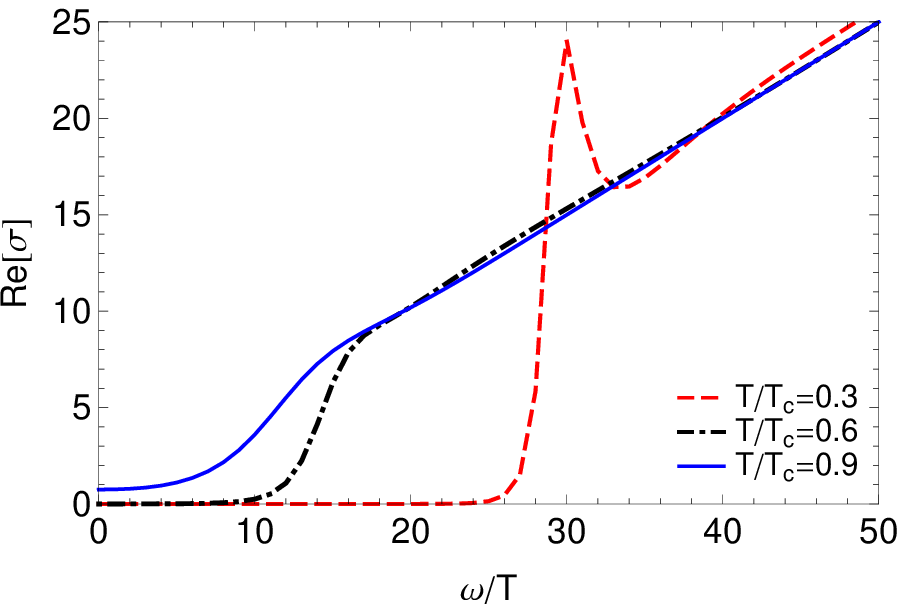}} \qquad %
\subfigure[~$z=2$, $m^{2}=0$, $b=0$]{\includegraphics[width=0.4\textwidth]{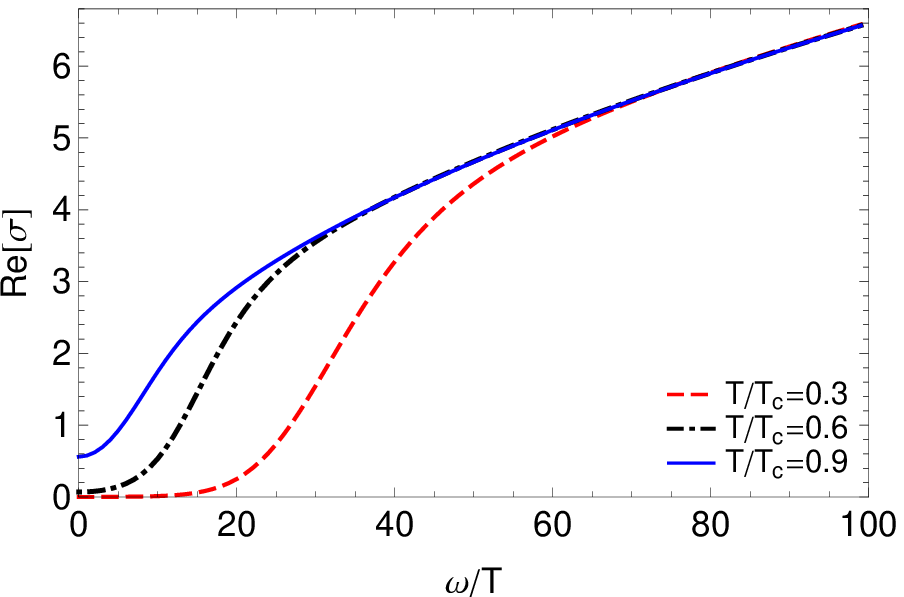}} \qquad %
\subfigure[~$z=2$, $m^{2}=0$, $b=0.04$]{\includegraphics[width=0.4\textwidth]{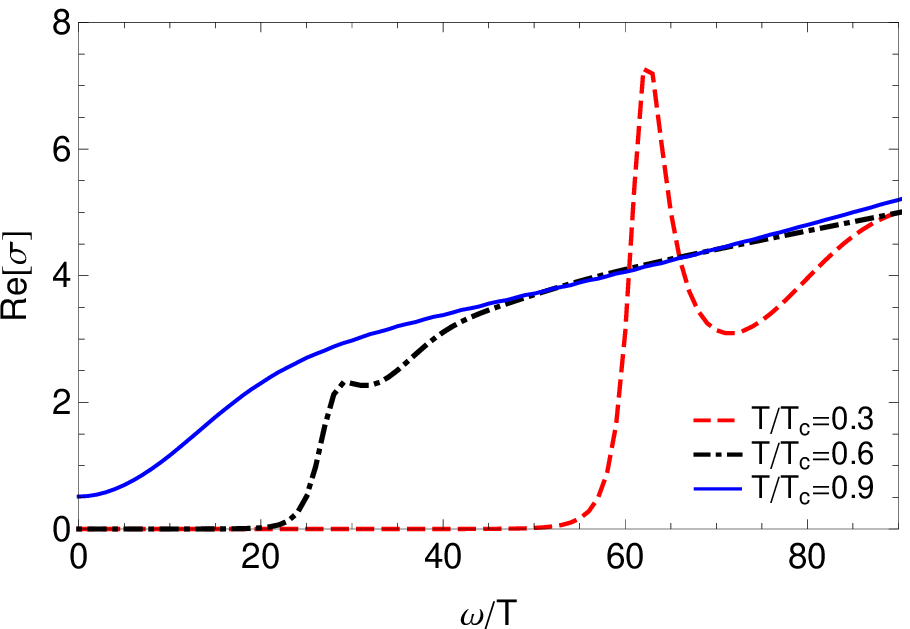}} \qquad %
\subfigure[~$z=1$, $m^{2}=-3/4$, $b=0$]{\includegraphics[width=0.4\textwidth]{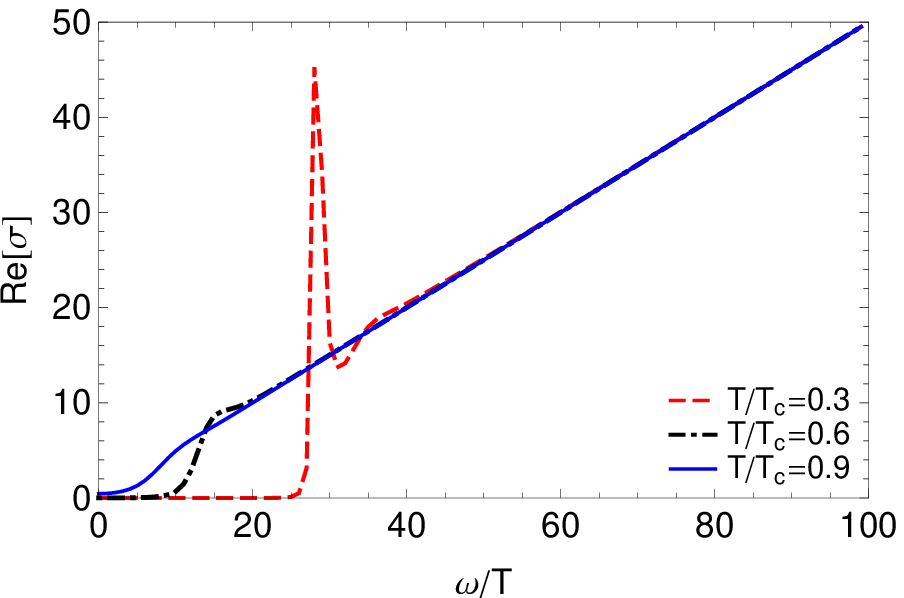}} \qquad %
\subfigure[~$z=1$, $m^{2}=-3/4$, $b=0.04$]{\includegraphics[width=0.4\textwidth]{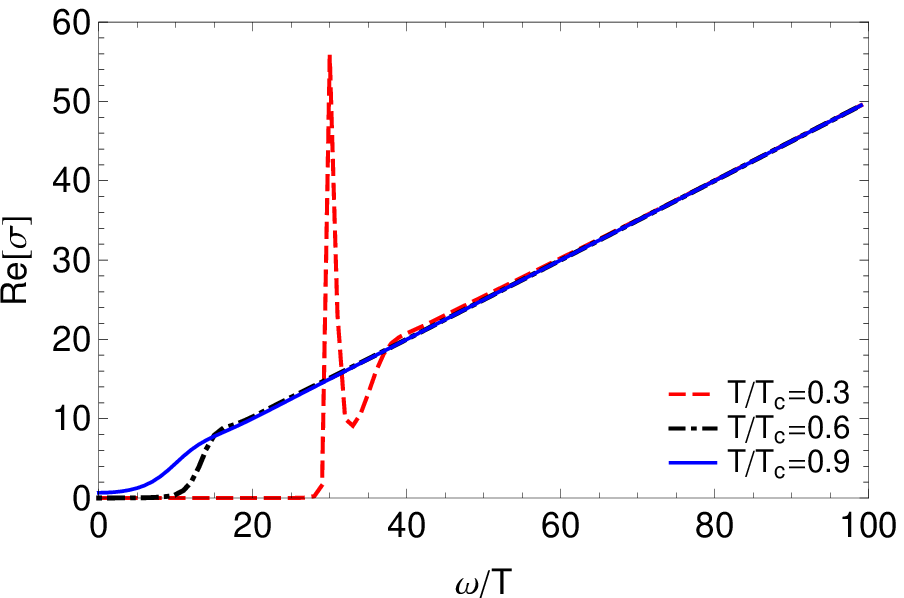}} \qquad %
\subfigure[~$z=2$, $m^{2}=-2$, $b=0$]{\includegraphics[width=0.4\textwidth]{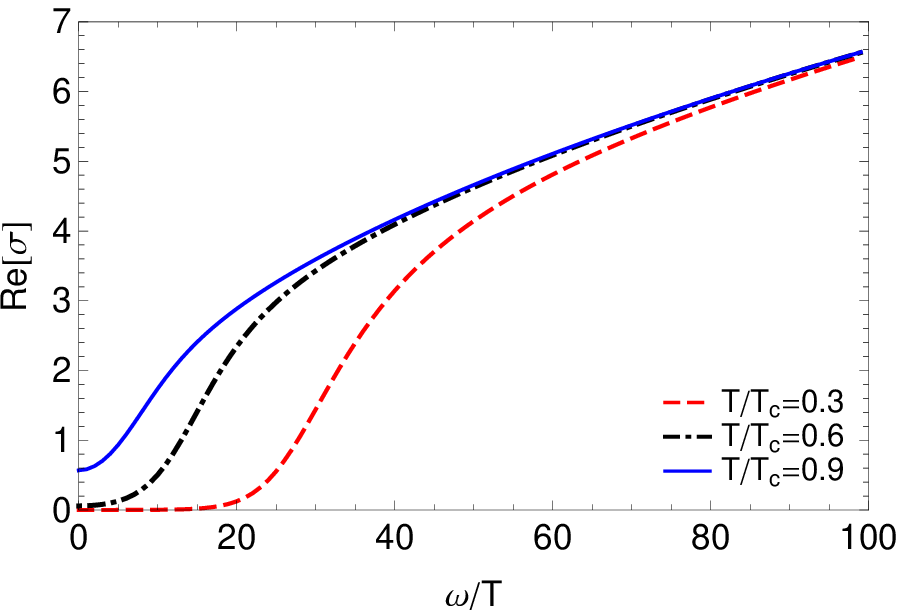}} \qquad %
\subfigure[~$z=2$, $m^{2}=-2$, $b=0.04$]{\includegraphics[width=0.4\textwidth]{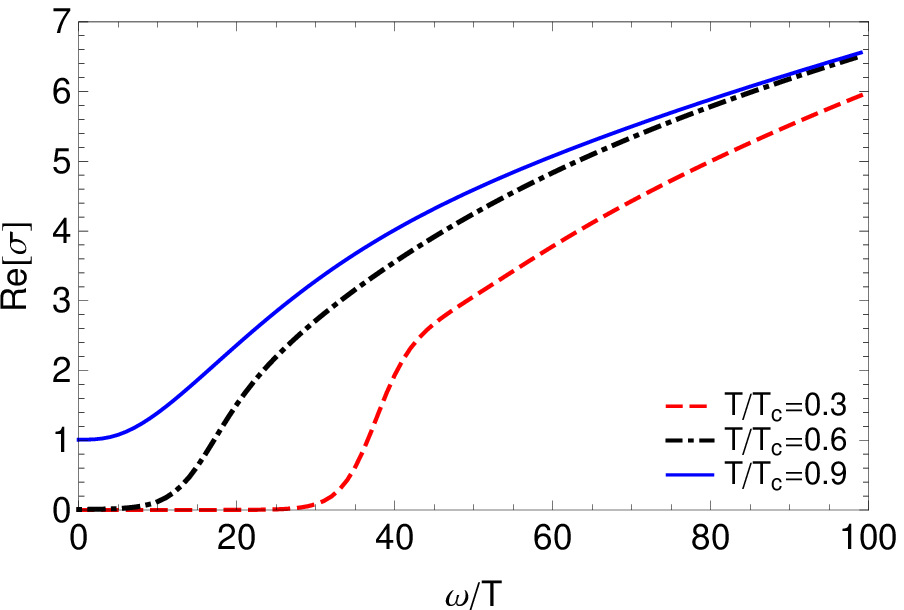}} \qquad %
\caption{The behavior of the real part of conductivity in $D=5$.}
\label{fig6}
\end{figure*}

\begin{figure*}[t]
\centering
\subfigure[~$z=1$, $m^{2}=0$, $b=0$]{\includegraphics[width=0.4\textwidth]{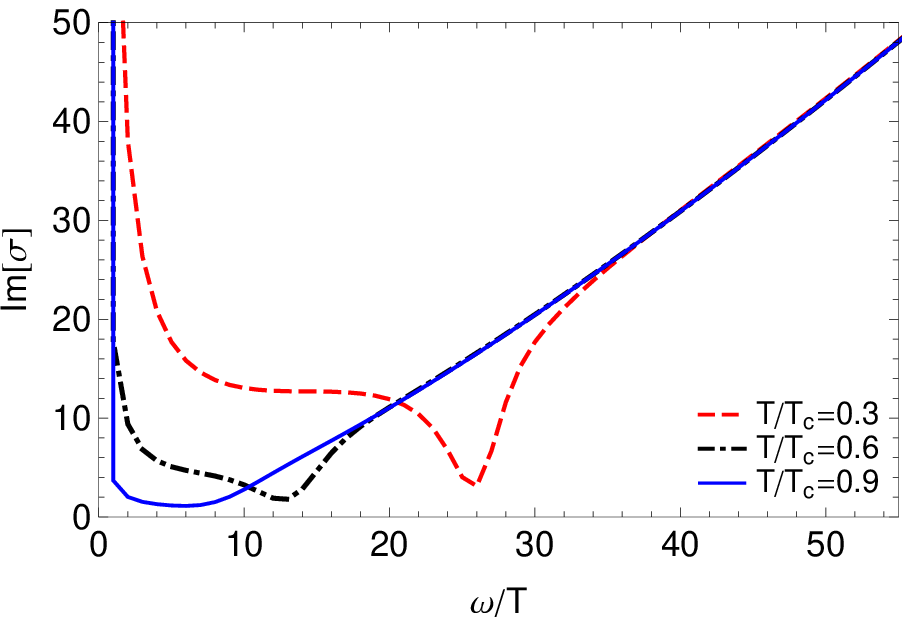}} \qquad %
\subfigure[~$z=1$, $m^{2}=0$, $b=0.04$]{\includegraphics[width=0.4\textwidth]{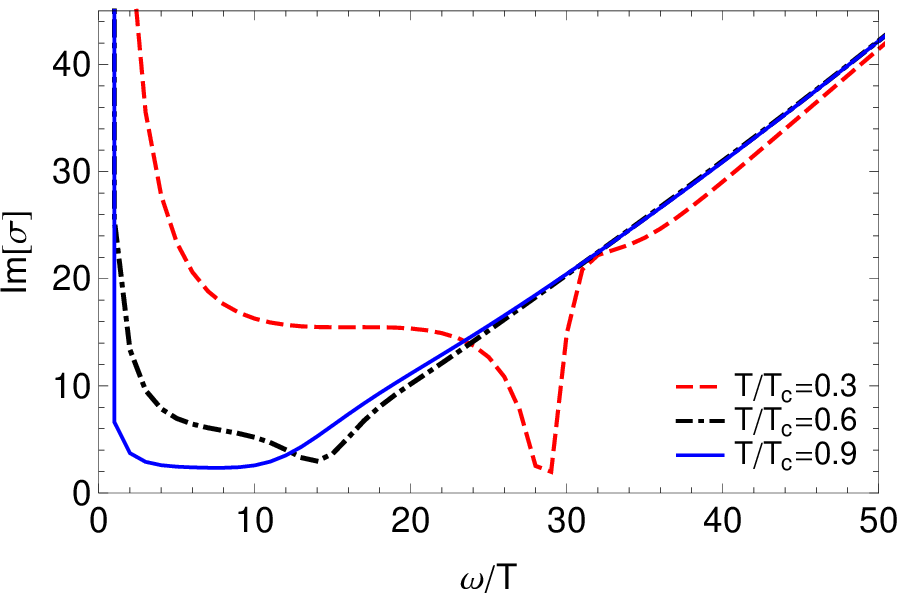}} \qquad %
\subfigure[~$z=2$, $m^{2}=0$, $b=0$]{\includegraphics[width=0.4\textwidth]{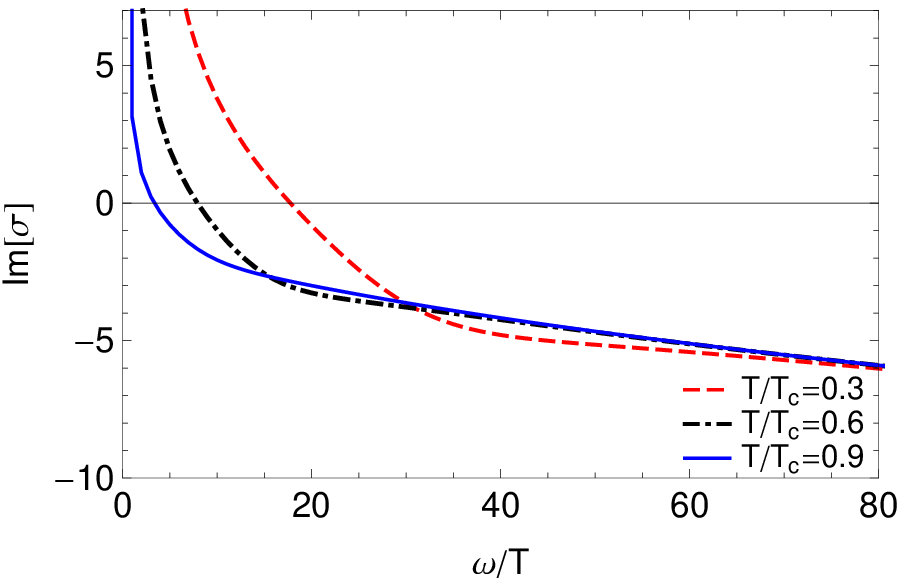}} \qquad %
\subfigure[~$z=2$, $m^{2}=0$, $b=0.04$]{\includegraphics[width=0.4\textwidth]{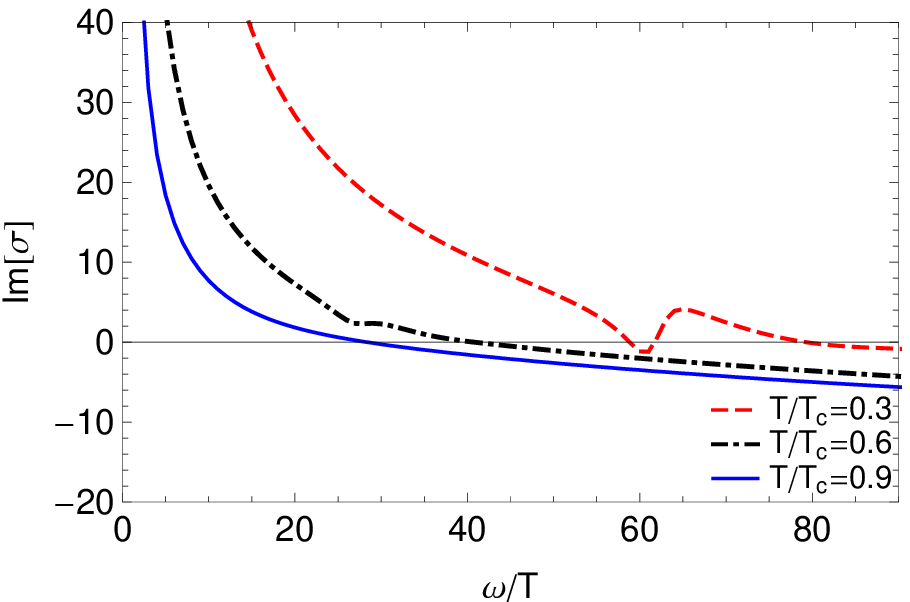}} \qquad %
\subfigure[~$z=1$, $m^{2}=-3/4$, $b=0$]{\includegraphics[width=0.4\textwidth]{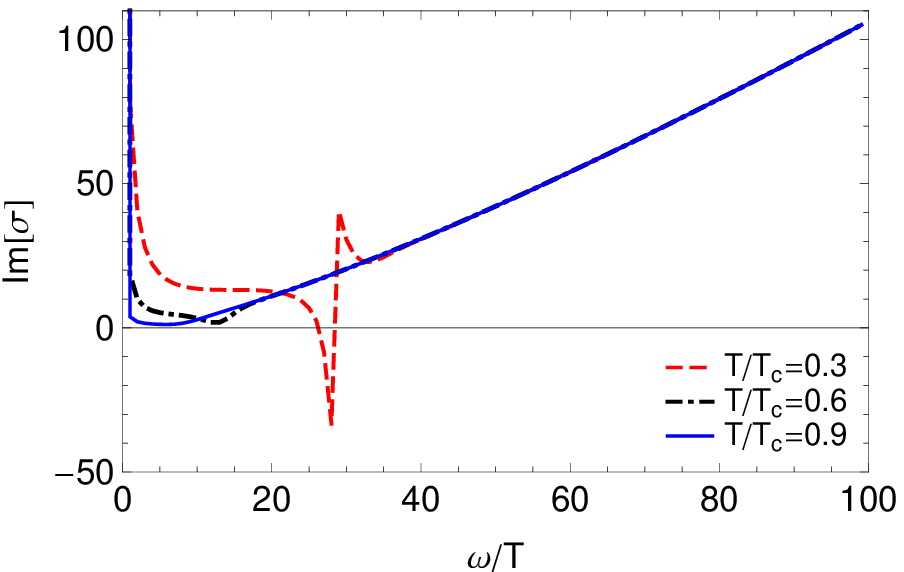}} \qquad %
\subfigure[~$z=1$, $m^{2}=-3/4$, $b=0.04$]{\includegraphics[width=0.4\textwidth]{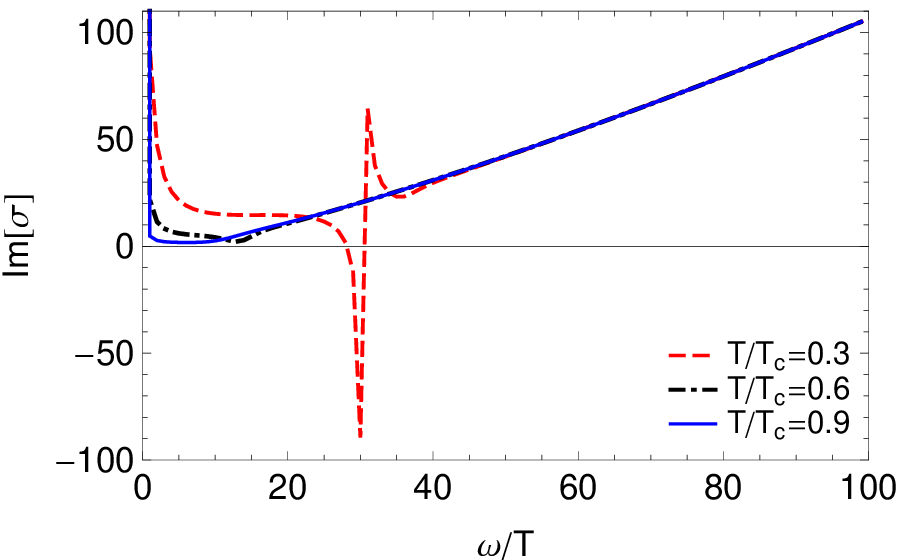}} \qquad %
\subfigure[~$z=2$, $m^{2}=-2$, $b=0$]{\includegraphics[width=0.4\textwidth]{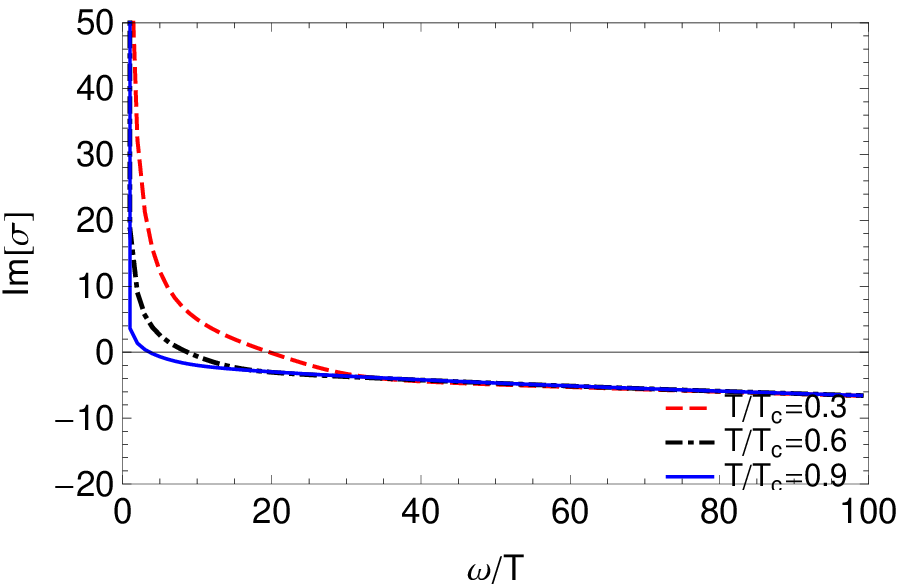}} \qquad %
\subfigure[~$z=2$, $m^{2}=-2$, $b=0.04$]{\includegraphics[width=0.4\textwidth]{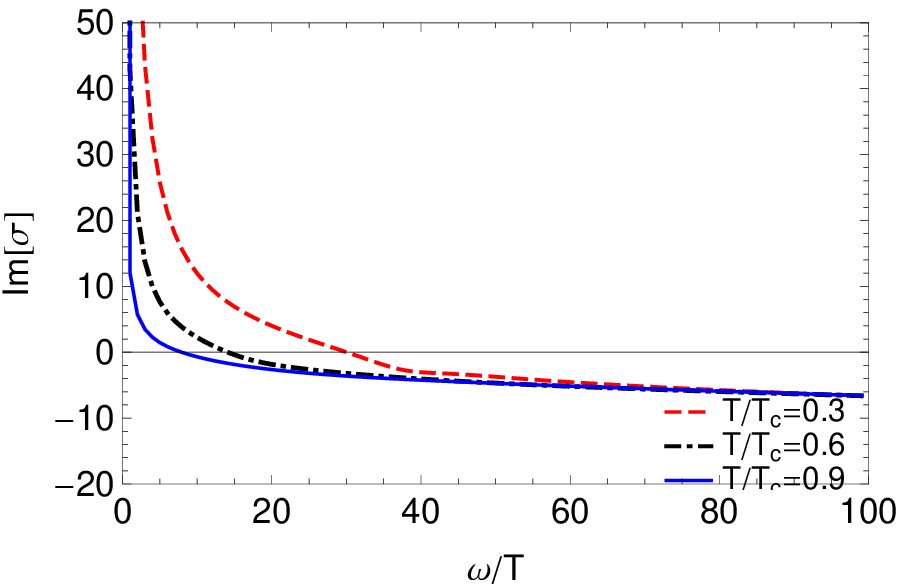}} \qquad %
\caption{The behavior of the imaginary part of conductivity in
$D=5$.} \label{fig7}
\end{figure*}

\begin{figure*}[t]
\centering
\subfigure[~$z=1$, $m^{2}=0$]{\includegraphics[width=0.4\textwidth]{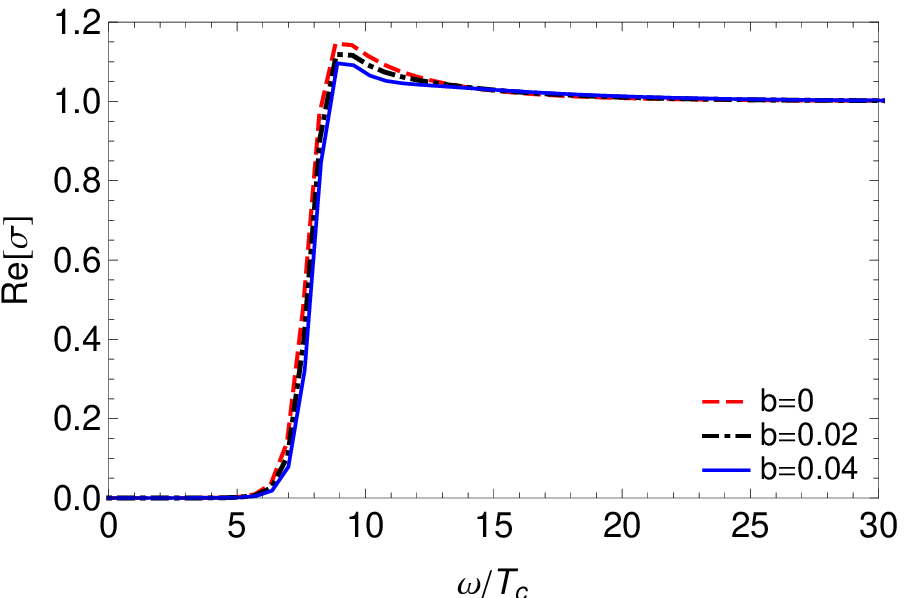}} \qquad %
\subfigure[~$z=2$, $m^{2}=0$]{\includegraphics[width=0.4\textwidth]{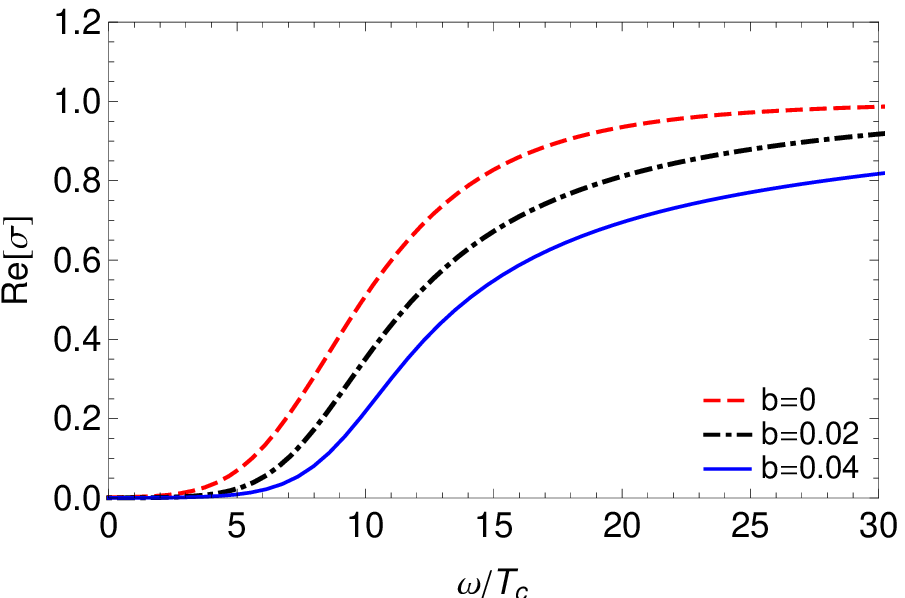}} \qquad %
\subfigure[~$z=1$, $m^{2}=3/4$]{\includegraphics[width=0.4\textwidth]{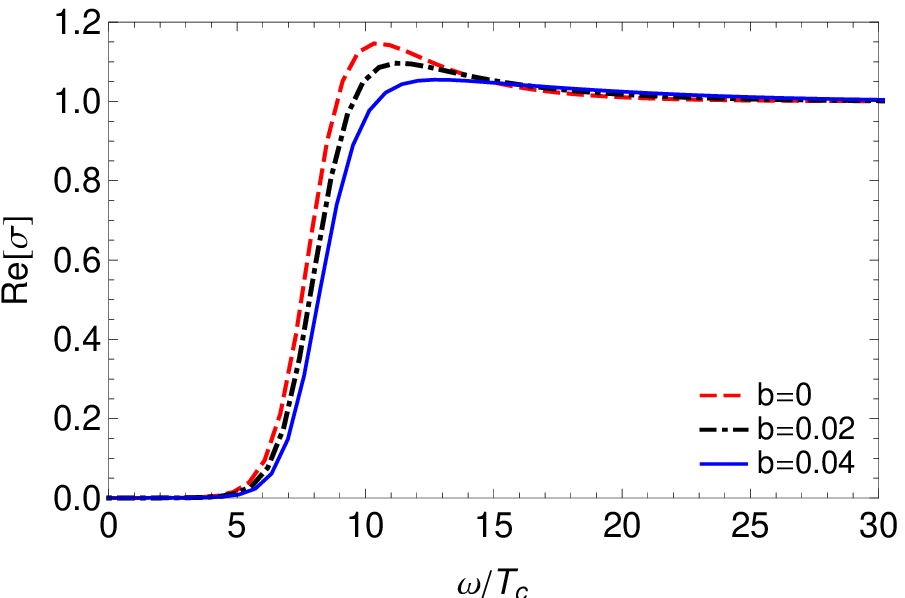}} \qquad %
\subfigure[~$z=2$, $m^{2}=-3/4$]{\includegraphics[width=0.4\textwidth]{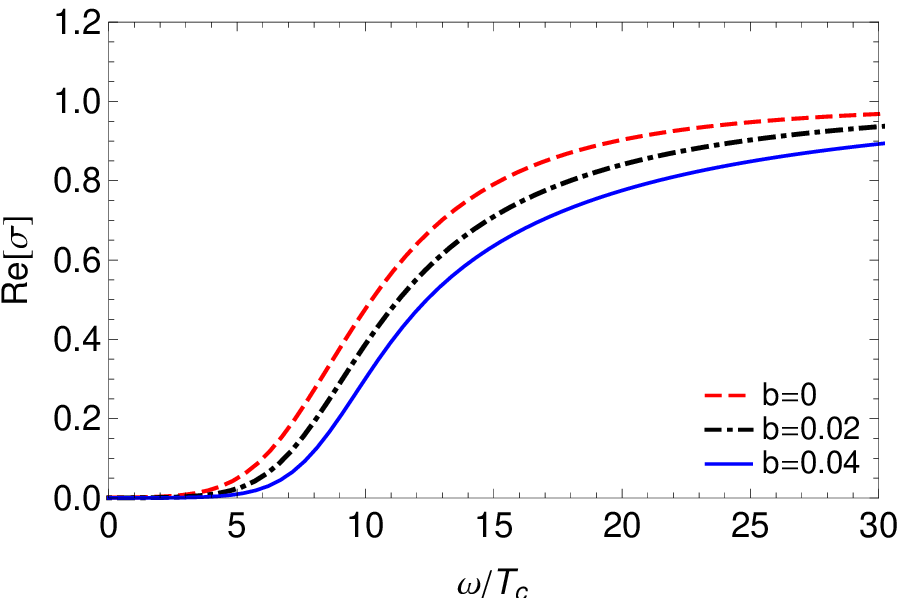}} \qquad %
\subfigure[~$z=1$, $m^{2}=0$]{\includegraphics[width=0.4\textwidth]{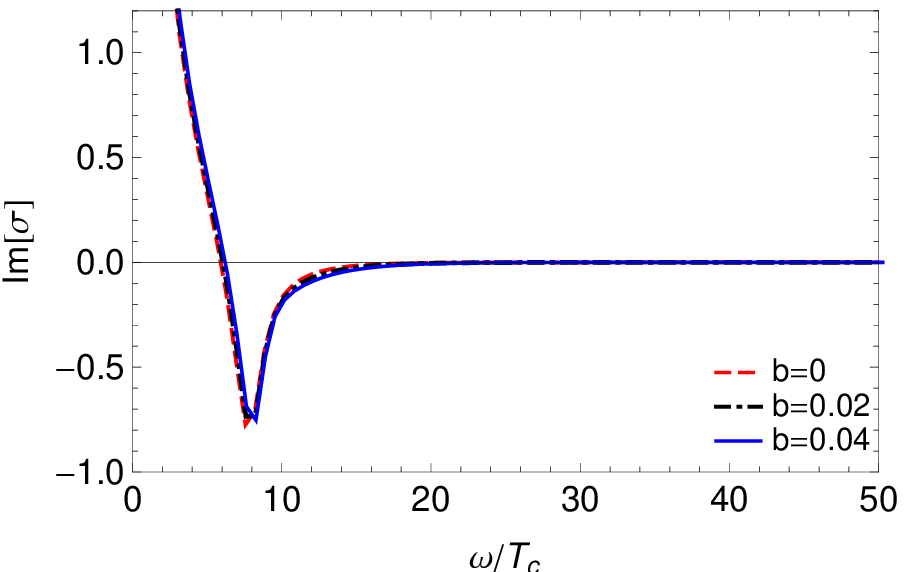}} \qquad %
\subfigure[~$z=2$, $m^{2}=0$]{\includegraphics[width=0.4\textwidth]{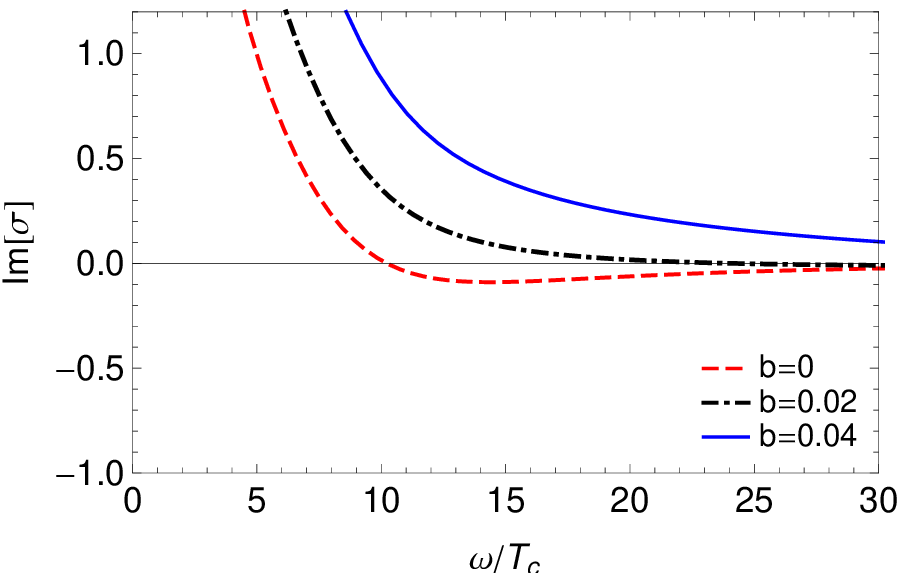}} \qquad %
\subfigure[~$z=1$, $m^{2}=3/4$]{\includegraphics[width=0.4\textwidth]{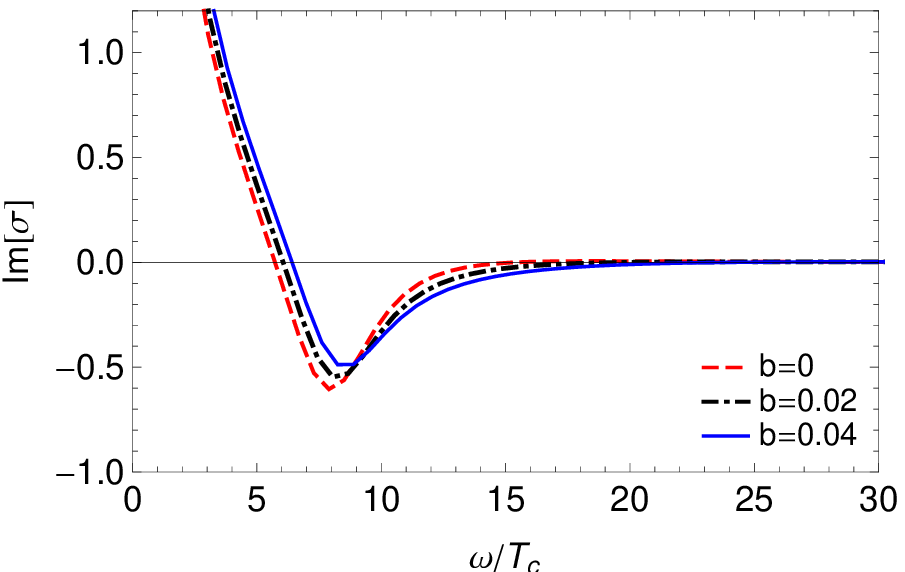}} \qquad %
\subfigure[~$z=2$, $m^{2}=-3/4$]{\includegraphics[width=0.4\textwidth]{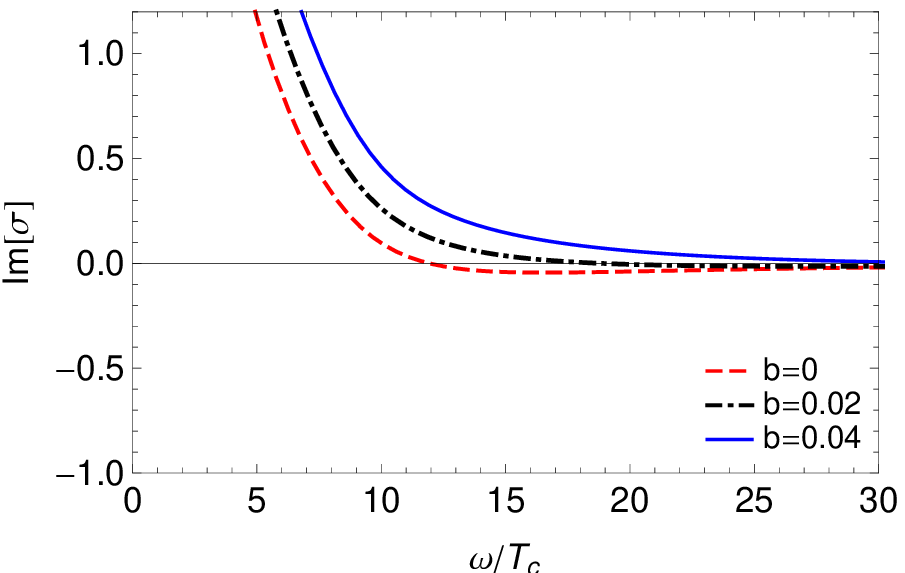}} \qquad %
\caption{The behavior of the real and imaginary parts of
conductivity in $D=4$ for $T/T_{c}=0.3$.} \label{fig10}
\end{figure*}

\begin{figure*}[t]
\centering
\subfigure[~$z=1$, $m^{2}=0$]{\includegraphics[width=0.4\textwidth]{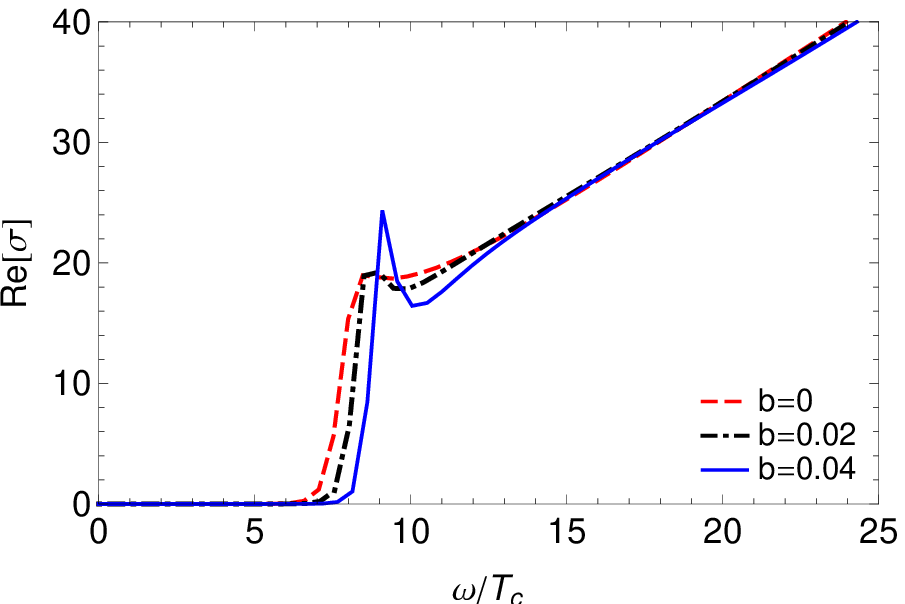}} \qquad %
\subfigure[~$z=2$, $m^{2}=0$]{\includegraphics[width=0.4\textwidth]{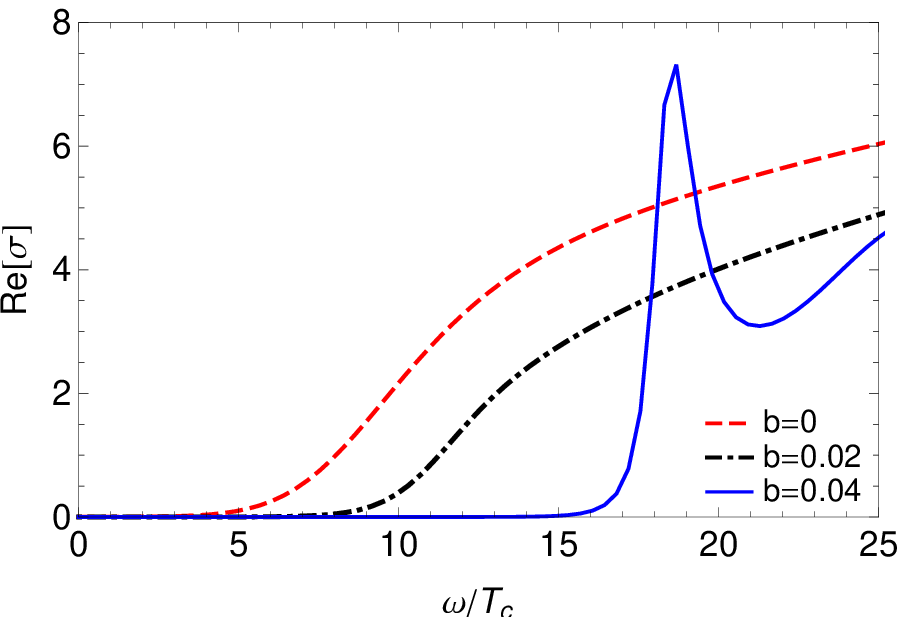}} \qquad %
\subfigure[~$z=1$, $m^{2}=-3/4$]{\includegraphics[width=0.4\textwidth]{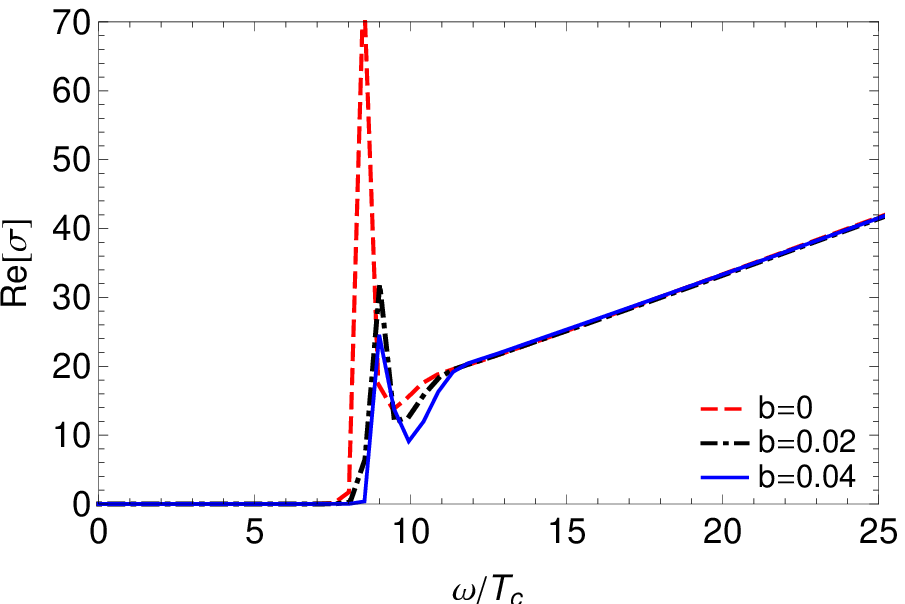}} \qquad %
\subfigure[~$z=2$, $m^{2}=-2$]{\includegraphics[width=0.4\textwidth]{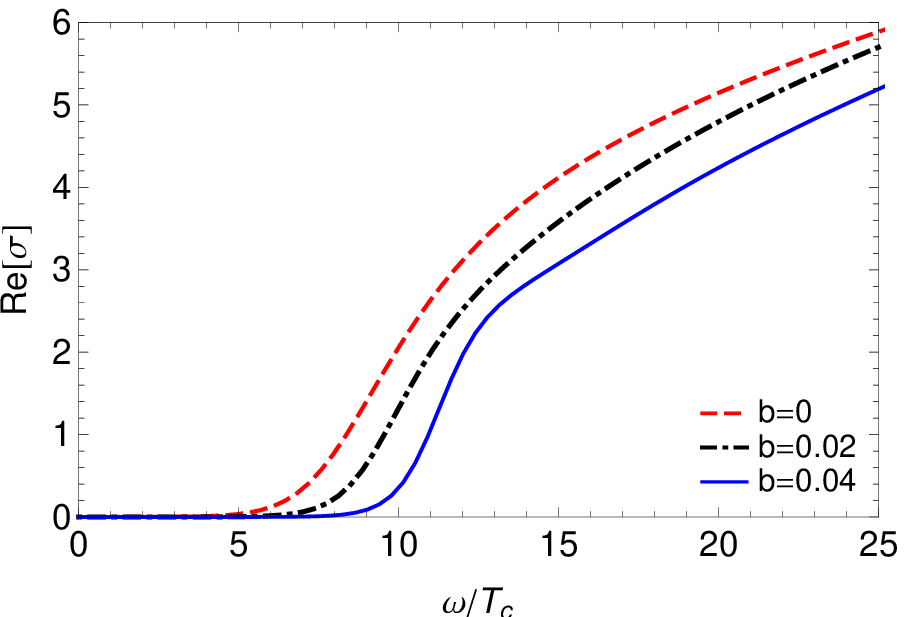}} \qquad %
\subfigure[~$z=1$, $m^{2}=0$]{\includegraphics[width=0.4\textwidth]{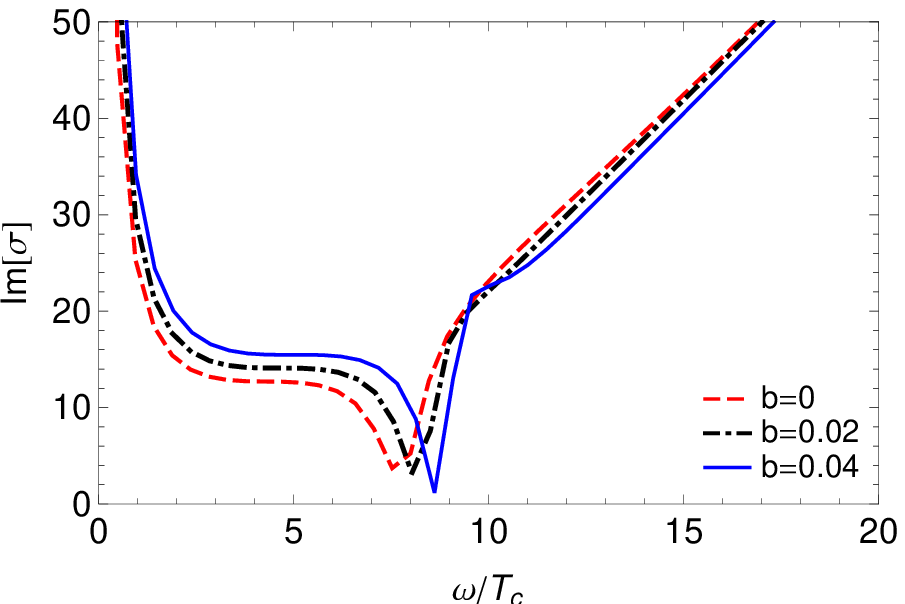}} \qquad %
\subfigure[~$z=2$, $m^{2}=0$]{\includegraphics[width=0.4\textwidth]{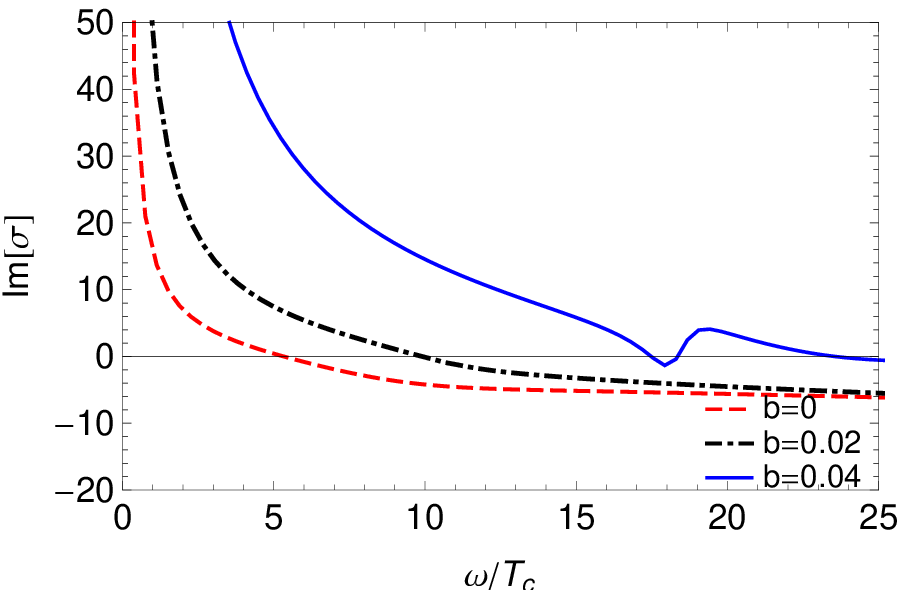}} \qquad %
\subfigure[~$z=1$, $m^{2}=-3/4$]{\includegraphics[width=0.4\textwidth]{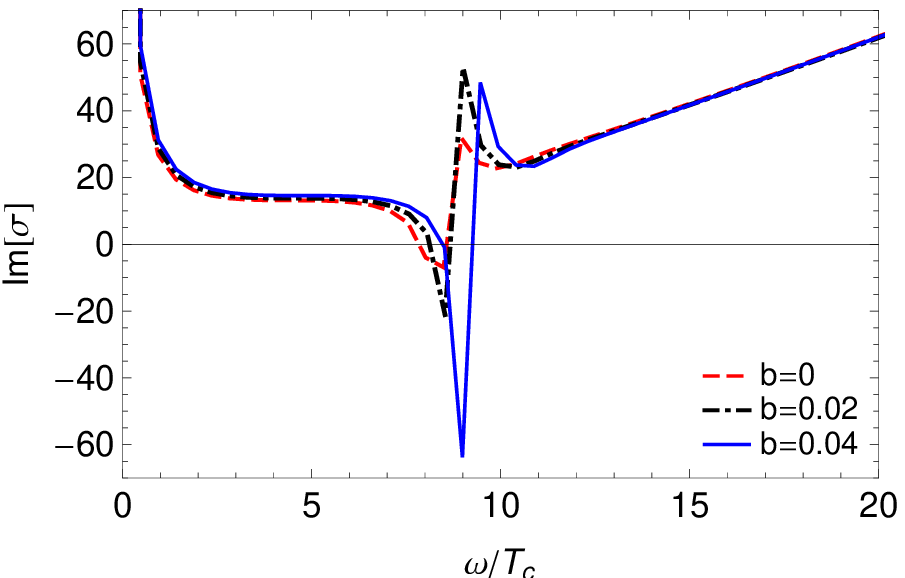}} \qquad %
\subfigure[~$z=2$, $m^{2}=-2$]{\includegraphics[width=0.4\textwidth]{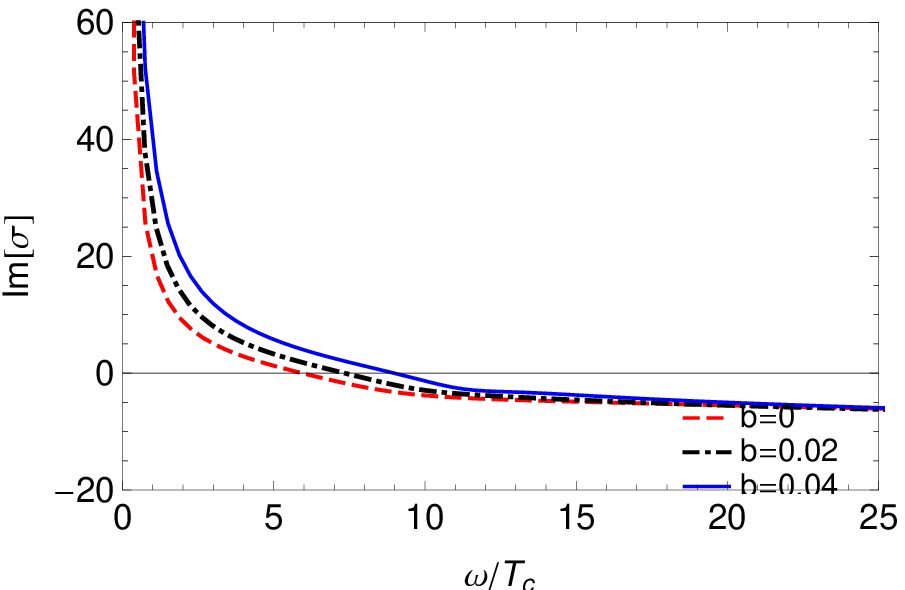}} \qquad %
\caption{The behavior of real and imaginary parts of conductivity
in $D=5$ for $T/T_{c}=0.3$.} \label{fig11}
\end{figure*}

In this section, we are going to investigate the effects of
Lifshitz scaling and nonlinear parameter on the electric
conductivity of holographic $p$-wave superconductor. For this
purpose, we apply appropriate electromagnetic perturbation by
turning on $\delta A_{y}=A_{y} e^{-i \omega t}$ on the black hole
background which acts as boundary electrical current in
holographic setup \cite{LU}. So, we have
\begin{equation} \label{eqcon}
A_y''(r)+ \left[\frac{d+z-1}{r}+\frac{f'(r)}{f(r)}+\frac{b r (z-1)
\phi '^2(r)}{b r^2 \phi '^2(r)-r^{2 z}}+\frac{b r^2 \phi '(r) \phi
''(r)}{r^{2 z}-b r^2 \phi'^2(r)}\right]A_y'(r)+ \left[\frac{\omega
^2}{r^{2 z+2} f^2(r) }-\frac{2 q^2 \rho_{x}^2(r) }{r^4
f(r)}\sqrt{1-\frac{b \phi '^2(r)}{r^{2 z-2}}}\right]A_y(r)=0.
\end{equation}
In the Maxwell limit Eq. (\ref{eqcon}), except a factor $2$ in the
last term which is originated from different approaches to
calculate conductivity, turns to corresponding equation in Ref.
\cite{LU}. The above equation has the asymptotic behavior as
\begin{equation}\label{eqaasym}
A_y''(r)+\frac{(d-1+z) }{r}A_y'(r)+\frac{\omega ^2}{r^{2 z+2}} A_y(r)=0,
\end{equation}
which admits the following solution for $A_{y}$,
\begin{equation} \label{aysol}
A_{y} = \left\{
\begin{array}{lr}
A^{(0)}+\frac{A^{(1)}}{r}+\cdots, & d=2, z=1\\
\bigskip\\
A^{(0)}+\frac{A^{(1)}}{r^{2}}+\cdots, & d=2, z=2\\
\bigskip\\
A^{(0)}+\frac{A^{(1)}}{r^2}+\frac{A^{(0)} \omega ^2 \log (\Omega r)}{2 r^2}+\cdots, & d=3,z=1\\
\bigskip\\
A^{(0)}+\frac{A^{(1)}}{r^{3}}+\cdots, & d=3, z=2\\
\end{array} \right.
\end{equation}%
where $A^{(0)}$, $A^{(1)}$ and $\Omega$ are constant parameters.
Furthermore, by considering $z=1$, Eqs. (\ref{eqaasym}) and
(\ref{aysol}) have the same form as in AdS case
\cite{mahyagaussp}. Based on gauge/gravity duality, the electrical
current is given by
\begin{equation} \label{eqj}
J=\frac{\text{$\delta $S}_{\text{bulk}}}{\text{$\delta
$A}^{(0)}}=\frac{\text{$\delta $S}_{o.s}} {\text{$\delta
$A}^{(0)}}=\frac{\partial(\sqrt{-g}\mathcal{L}_{m})}{\partial
A_y'}\vert r\rightarrow \infty,
\end{equation}
in which the on-shell bulk action $S_{o.s}$ by using equation (\ref{eqj}) is defined by
\begin{equation}\label{sos}
S_{o.s.}=\int_{r_+}^{\infty }\, dr \int \, {d}^{d-1}x
\sqrt{-g}\mathcal{L}_{m}=- \frac{1}{2} \int \, {d}^{d-1}x
\left[\frac{ f(r) A_{y}(r) A_{y}'(r)}{1-b r^{2-2 z} \phi
'^2(r)}\right] r^{d+z-1} \sqrt{1-b r^{2-2 z} \phi '^2(r)}.
\end{equation}
The electrical conductivity in a corresponding framework is
\cite{H08}
\begin{equation}\label{eqohm}
\sigma_{yy}=\frac{J_{y}}{E_{y}}, \  \  \  E_{y}=-\partial_{t} \delta A_{y}.
\end{equation}
Employing Eqs. (\ref{eqj}), (\ref{sos}) and (\ref{eqohm}) and
using appropriate counterterms, based on the re-normalization
method to remove the divergency, the electrical conductivity is
obtained as \cite{skenderis}
\begin{equation} \label{sigma}
\sigma_{yy} = \left\{
\begin{array}{lr}
\frac{A^{(1)}}{i \omega A^{(0)} }, & d=2, z=1\\
\bigskip\\
\frac{2 A^{(1)}}{i \omega A^{(0)} }, & d=2, z=2\\
\bigskip\\
\frac{2 A^{(1)}}{i \omega A^{(0)} }+\frac{i}{2}\omega, & d=3,z=1\\
\bigskip\\
\frac{3 A^{(1)}}{i \omega A^{(0)} }, & d=3, z=2\\
\end{array} \right.
\end{equation}%
For $z=1$, we obtain the same equations as in the AdS background
\cite{mahyagaussp}. In order to follow our research, we imply an
ingoing wave boundary condition near the horizon as
\begin{equation}\label{ayexpand}
A_{y}(r)=f(r)^{-i \omega/ (4 \pi T)} \left[1+a(1-r)+b(1-r)+\cdots\right],
\end{equation}
where by Taylor expansion of equation (\ref{eqaasym}) around the
horizon, coefficients $a$ and $b$ are obtained. The behavior of
real and imaginary parts of conductivity as a function of $\omega
/ T$ are shown in Figs. \ref{fig4}-\ref{fig7}. The conductivity
along the $y$ direction in Lifshitz holographic $p$-wave
superconductors is the same as $\sigma_{xx}$ in $s$-types
\cite{LU}. Although the figures follow different trends but in all
cases the behavior of real part of conductivity in large frequency
regime can be predicted by a power law function as
$Re[\sigma]=\omega^{D-4}$ similar to \cite{mahyagaussp}. The real
and imaginary parts of conductivity follow the Kramers-Kronig
relation. Thus, we observe the appearance of the delta function
and pole, respectively. By increasing the Lifshitz scalaing $z$,
the gap energy becomes unclear like the minimum of the imaginary
part. However, in some cases, we observe that by decreasing the
temperature \cite{Bu} for strong BI nonlinear electrodynamics, the
gap and minimum in real and imaginary parts of conductivity appear
which is obvious in $D=5$ with $m^{2}=0$ and $b=0.04$. When $z=1$,
below the critical temperature, the superconducting gap appears
and becomes deeper and sharper by diminishing the temperature
which yields to larger values of $\omega_{g}$ which makes the
conductor/superconductor phase transition harder to form because
it can be interpreted as the energy needs to break the fermion
pairs. With the use of figures \ref{fig10}-\ref{fig11} which are
graphed in $T=0.3 T_{c}$, the value $\omega_{g}=8 T_{c}$ is
generally achieved for $z=1$ which differs from the predicted
value $3.5$ of BCS theory. This difference is originated from the
fact that holographic superconductors are strongly coupled systems
and because of this character, they are expected to be suitable
for description the high temperature superconductors \cite{caipp}.
In addition, larger values of nonlinearity parameter shifts the
maximum and minimum parts of conductivity toward larger values but
the effects of nonlinearity on the value of energy gap depends on
our choice of mass which implies the fact that the value of
$\omega_{g}/T_{c}$ is characterized by our selection of mass $m$,
nonlinearity $b$ and dynamical critical exponent $z$ in each
dimension.

\section{summary and conclusion}\label{section3}
In this work by employing the gauge/gravity duality, we have
studied the holographic $p$-wave superconductors with
Lifshitz scaling in the presence of BI nonlinear electrodynamics.
We applied the shooting method to calculate the
equations of motion and analyze the behavior of the condensation
as a function of temperature numerically. We found the relation between
critical temperature $T_{c}$ and $\rho^{z/d}$ for different values
of mass $m$, nonlinearity effect $b$ and Lifshitz scaling $z$ in
$4D$ and $5D$ spacetime. Based on our results, we observe that the
temperature decreases with increasing each of three parameters
$m$, $z$ and $b$, which means that superconductivity phase faces
with more difficulties to occur. The condensation behavior in
Lifshitz scaling is similar to AdS spacetime by obeying the mean
field trend in the vicinity of critical point. Increasing the
anisotropy between space and time, diminishes the condensation
value. After that, by applying a suitable perturbation on black
hole background as $\delta A_{y}=A_{y} e^{-i \omega t}$, we
investigated the effects of Lifshitz scaling on the electrical
conductivity of the holographic $p$-wave superconductors and plot
the behavior of real and imaginary parts of conductivity as a
function of frequency. The plotted Figures are different with each
other but they follow some universal behaviors. For instance, in
large frequencies, we can predict the behavior of real part of
conductivity as $Re[\sigma]=\omega^{D-4}$. In addition, the real
and imaginary parts of conductivity are related to each other via
the Kramers-Kronig relation. Actually, the real part shows a delta
function behavior and the imaginary part has a pole at zero
frequency. At low frequencies with $z=1$, real and imaginary parts
of conductivity show a gap energy and minimum which shift toward
larger frequencies by diminishing temperature. In this regime,
increasing the Lifshitz critical exponent $z$, makes the gap
energy and minimum unclear. However in some cases, they were
occurred by going down temperatures and raising the nonlinearity
effect. Our choice of mass in each dimension has a direct outcome
on the effect of BI nonlinear electrodynamics on conductivity but
generally the gap energy and minimum of conductivity shift toward
larger values of frequency by enlarging the nonlinearity effect.
In the limiting case where $z=1$, the ratio
$\omega_{g}/T_{c}\simeq 8$ is obtained generally which is larger
that the BCS value because of the strong coupling between the
pairs.
\begin{acknowledgments}
We are deeply grateful to S.A. Hartnoll for his helpful numerical
code which was published for public use. We appreciate the
Research Council of Shiraz University. A.S. thanks Hermann Nicolai
and Max-Planck-Institute for Gravitational Physics (AEI), for
hospitality.
\end{acknowledgments}


\end{document}